\documentclass[12pt]{article}
\usepackage{tikz}
\usepackage[utf8]{inputenc}
\usepackage{amsthm}
\usepackage{hyperref}
\usepackage{amssymb,amsmath,latexsym,wrapfig}
\usepackage{graphicx}
\usepackage{psfrag}
\usepackage[top=3cm, bottom=2.5cm, left=3cm, right=2cm]{geometry}
\begin{document}
\title{{Stable super-inflating cosmological solutions in f(R)-gravity}}
\author{Mikhail M. Ivanov,\!$^{a}$\footnote{E-mail: mm.ivanov@physics.msu.ru}  Alexey V. Toporensky,\!$^{b}$\footnote{E-mail: atopor@rambler.ru}\vspace{.2cm}\\
\normalsize\llap{$^a$} \it Faculty of Physics, Moscow State University, \\
      \normalsize \it  Leninskiye gory 1-2, Moscow 119991, Russia\\
\normalsize\llap{$^b$}
 \it Sternberg Astronomical Institute, Moscow State University,\\
 \normalsize\it Universitetsky prospect, 13, Moscow 119992, Russia
}
\date{}
\maketitle
\begin{abstract}

We consider super-inflating solutions in modified gravity for several popular families of $f(R) $ functions. Using scalar field
reformulation of $f(R)$-gravity we describe how the form of effective scalar field potential can be used for
explaining  existence of stable super-inflation solutions in the theory under consideration. Several new solutions of this type
have been found analytically and checked numerically.
\end{abstract}
\section{Introduction}
Theories of modified gravity first introduced in order to match General relativity
and quantum physics, became very popular recently after discovery of accelerated
expansion of our Universe. Being higher order, equation of motion of these theories
can contain some regimes impossible in General relativity with its 2-d order
equations. Some of new regimes may be useful for explaining observed accelerated
expansion, others can spoil standard cosmology, so constructing viable models
appears to be rather tricky (however, not impossible).

On the other hand, though modification of an early Universe dynamics gives
one of 
the earliest proposed inflation scenario (so called Starobinsky inflation,\cite{Starobinsky}), 
it is less popular now due to success of scalar field inflation theory 
\cite{Linde},\cite{Lyth}. This theory does not
require any modification of General Relativity, so any correction to high-energy
dynamics being pre-inflational have no consequences today. However, this statement
is valid only if such corrections do not modify significantly inflation itself. In particular,
stable super-inflating vacuum solutions (solutions with increasing Hubble parameter)
are dangerous. Recent analysis shows that stable phantom solutions are common
in $f(R)$ gravity with power-law $f(R)$ \cite{CDCT},\cite{CTD}. This regime leads to Big Rip singularity
in a finite time interval \cite{BigR}\cite{Grave3},
and can not be followed by some other non-singular regime.

Apart from the phantom, other super-inflating solution in $R+\alpha R^2$ gravity in known for
decades. It have been discovered initially in \cite{Ruzm} and rediscovered
later  (\cite{B&H}\cite{GL}). The main feature of this solution is that the Hubble parameter grows
linearly in time, so it takes infinite time before it reaches infinity. In generalized
modified gravity other forms of this solution are possible, in particular $R\Box R$ corrections
in the action lead to  $H \sim t^{1/3}$ solution \cite{STT}.
Moreover, the presence of matter with some exotic equations of state or rather exotic functions $f(R)$ 
leads to super-inflation with $H \sim e^t$ \cite{LR},\cite{LR1},\cite{LR2},\cite{LR3},\cite{Yurov} in these papers such solution
is called as "Little Rip".
However, from physical point of view, Ruzmaikin solution with its analogs and phantom solutions are not
substantially different - both lead to Planck density reached in finite time interval, and
prolongation of the solution further in time has no any physical justification.
Both solution, if realised in early epoch of Universe evolution can precede inflation
only being unstable. It should be noted that recently some models describing "graceful exit"
from a phantom regime have been constructed \cite{Yurov, Piao}, however, they are dealing with a phantom matter
as a source of this regime, and can not be directly applied to $f(R)$-cosmology. 

In the present paper we try to explain why stable super-inflating solution are so
natural in $f(R)$ gravity and present some of them which have not been found earlier.

\section{Effective scalar degree of freedom}

$f(R)$ gravity is known to be dynamically equivalent to the General Relativity with 
the minimally coupled scalar field (\cite{Sot},\cite{fR},\cite{MG},\cite{Sotiriou}). 
Therefore, we 
expect that properties of cosmological solutions in $f(R)$-gravity 
can be guessed from the behavior of effective scalar field potential.
We begin with the equations of motion and their trace \cite{Sotiriou} for 
$f(R)$-gravity in the Jordan frame
\footnote{The signature of the metric is assumed 
to be $(-,+,+,+)$ and $c=8\pi G=1$.}
.
\begin{align}
\label{eqm}
f'G_{\mu\nu}-f'_{;\mu\nu}+\Bigl[\square f'-\frac{1}{2}(f-f' R)\Bigr]g_{\mu\nu} =T_{\mu\nu}\;,\\
\square f' = \frac{1}{3}(2f-f'R)+\frac{T}{3}\,.
\end{align}
Following  \cite{Frolov}, we  introduce the scalar degree of freedom and define its potential:
\begin{align}
\label{p}
 \phi &\equiv f'-2\,,\\
 \label{V1}
  V'(\phi) &\equiv \frac{1}{3}(2f-f'R)\Rightarrow \\
  \label{V}
  \begin{split}
  V(\phi)&=\frac{1}{3}\int{(2f-f'R)\frac{d\phi(R)}{dR}dR}= \frac{1}{3}\int(2f-f'R)f''dR \,,
  \end{split}
  \\
   \frac{T}{3}&\equiv -\mathcal{F}\,.
\end{align}
Then the trace equation becomes:
\begin{align}
\square \phi & = V'(\phi)-\mathcal{F}\,,
\end{align}
Finally, in the flat FLRW metric ($ds^2=-dt^2+a(t)^2\mathbf{dx}^2$) the above equation is 
\begin{align}
\label{parteq}
\ddot{\phi}=-V'(\phi)+\mathcal{F}-3H\dot{\phi}\,.
\end{align}
After the substitution $\phi \longrightarrow x$ we obtain a simple one-dimensional 
equation of motion for a "particle" in the potential $V(x)$ with the friction 
$3H\dot{x}$ and the driving force $\mathcal{F}$.
This analogue is  useful because behavior of the mechanical  system is quite evident especially
in the absence of external force.
Using the Eqs.(\ref{V}),(\ref{p}) one can see that the function $V(\phi)$ is represented by a plane 
curve defined parametrically $[\phi(R),V(R)]$, where $R=R(t)$. 
We consider the vacuum case ($T=-3\mathcal{F}=0$) - this case 
corresponds to the absence of driving force.\par 
The equations of motion (\ref{eqm}) for FLRW metric have the form
\begin{align}
\label{numeq}
3f'H^2 &= \frac{1}{2}(f'R-f)-3H\dot{f'}\,,\\
\label{R}
R&=6(2H^2+\dot{H}).
\end{align}

In the present paper we consider mostly regular functions $f(R)$ which remain finite for finite $R$
with its first derivative\footnote{It should be also noted that for this theory to be perturbatively stable, one needs $f'>0$ to 
avoid ghosts (\cite{Nunez}) and $f''>0$ to avoid tachyon instability 
(\cite{Sawicki},\cite{Dolgov}).}. 
As we will see further, properties of the effective potential allow 
us to find some asymptotic cosmological solutions for particular 
functions $f(R)$, which appears to be useful  for detection of a 
possible super-inflating behavior.\par 

Movement in the plane [$\phi ,V$] corresponds to the solution $R(t)$. 
On the other hand, the composite function $\phi\bigl(R(t)\bigr)$ obeys 
the equation of motion for "particle" in classical mechanics (\ref{parteq}).
Thus let us consider the "particle" motion.
The following situations are possible:
\begin{enumerate}
\item $V(\phi)$ has a local minimum,
hence in the limit of late time the "particle" would fall down in this minimum 
due to the Hubble friction. 
Scalar degree of freedom takes the value $\phi=\phi_{a}$, 
for which $V'(\phi_{a})=0$. Therefore, $2f(R_a)-f'(R_a)R_a=0$ (see (\ref{V1})) 
and we see that $R_{a}=R(\phi_{a})$ corresponds to the stable de Sitter (dS) ($H=const$), 
or Minkowski ($H=0$) asymptotic solution.
\item $V(\phi)$ has neither local nor global minima, hence it is unbounded from 
below. The "particle"\, must fall into an unrestricted area of the potential curve. If $|V'(\phi)|$ is unbounded for $V \to -\infty$
this, according to Eq.(\ref{V1}) indicates an 
 infinite growth of $|R|$.
Stable super-inflating solutions in the signature chosen correspond to $R \to + \infty$
solutions. 
The case of limited $V(\phi)$ for infinite $R$ should be excluded from our considerations. We have not seen
it in particular families of $f(R)$ functions studied in the present paper (see below).
\end{enumerate}
If $V(\phi)$ has the local minimum (possibly more than one) and is unbounded 
from below then different asymptotic regimes are possible depending on initial 
conditions.

It should be noted that due to presence of non-standard singularities in $f(R)$ gravity (which occurs when a coefficient 
at the highest derivative term in the equations of motion vanishes) the cosmological evolution does not necessarily ends up at
some smooth asymptotic regime. That is why existence of such regimes can only be guessed from the form of effective potential
and should be checked explicitly. In the present paper we write down analytic from of super-inflating solutions in question
and check their stability only numerically, leaving full analytic treatment of stability (regarding solution which have not been
known earlier) for a future work.  
As for possible unstable solutions (which may be interesting from the viewpoint
of the Starobinsky inflation) the obvious example is a potential which has a local maximum. We will not consider unstable solutions 
further in this paper.\par 
It also should be noted that the proposed effective potential does not help us to study the regimes where
$f'(R)$ becomes zero at some finite $R$ and the effective gravitational constant changes sign. 
To be sure that a phase curve can safely cross this point, one should apply the commonly used conformal transformation to the Einstein frame (\cite{Sot},\cite{fR},\cite{MG},\cite{Sotiriou}). However, as the purpose of this work is to study super-inflating regimes in $f(R)$ gravity the method proposed in \cite{Frolov} provides relatively simple scheme of detecting a super-inflation behavior based on mechanical analogue because the dynamics of the effective scalar field in this parametrisation resembles particle motion with friction in the potential field. The standard way \cite{fR} may be useful in describing a global picture of corresponding dynamics
which is beyond the scope of the present paper.

\section{Examples of $f(R)$-functions and numerical results}

\subsection{f(R)=$R+\alpha R^N$} 
\begin{figure}[h!]
\begin{minipage}[h!]{0.35\linewidth}
\includegraphics[width=1\linewidth]{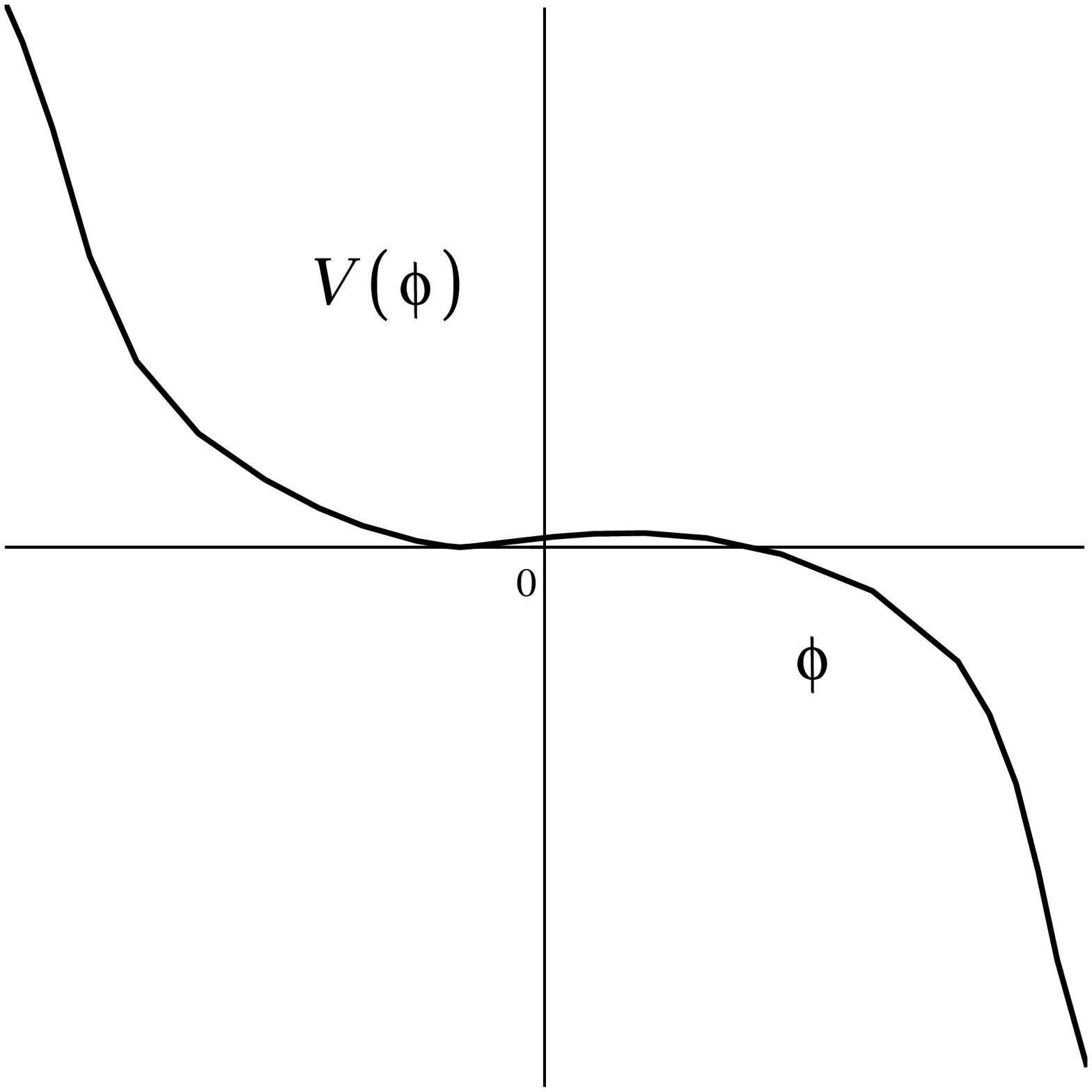}
\caption{Effective potential of scalaron for $f(R)=R+R^4$.}
\label{ris:R+R4}
\end{minipage}
\hfill
\begin{minipage}[h!]{0.35\linewidth}
\includegraphics[width=1\linewidth]{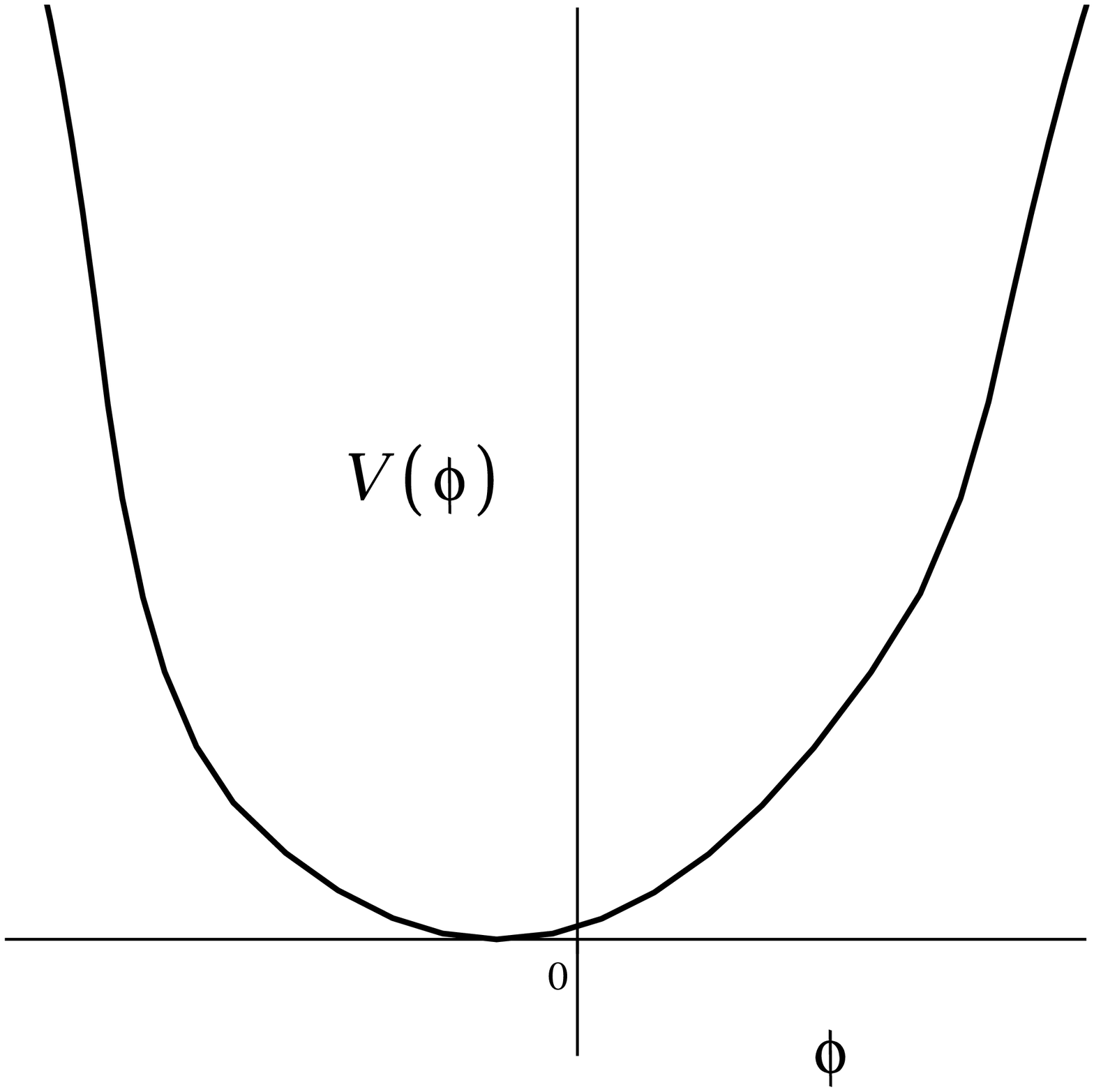}
\caption{Effective potential of scalaron for $f(R)=R+R^2$.} 
\label{ris:R+R2} 
\end{minipage}
\vfill
		\begin{minipage}[h!]{0.35\linewidth}
			\includegraphics[width=1\linewidth]{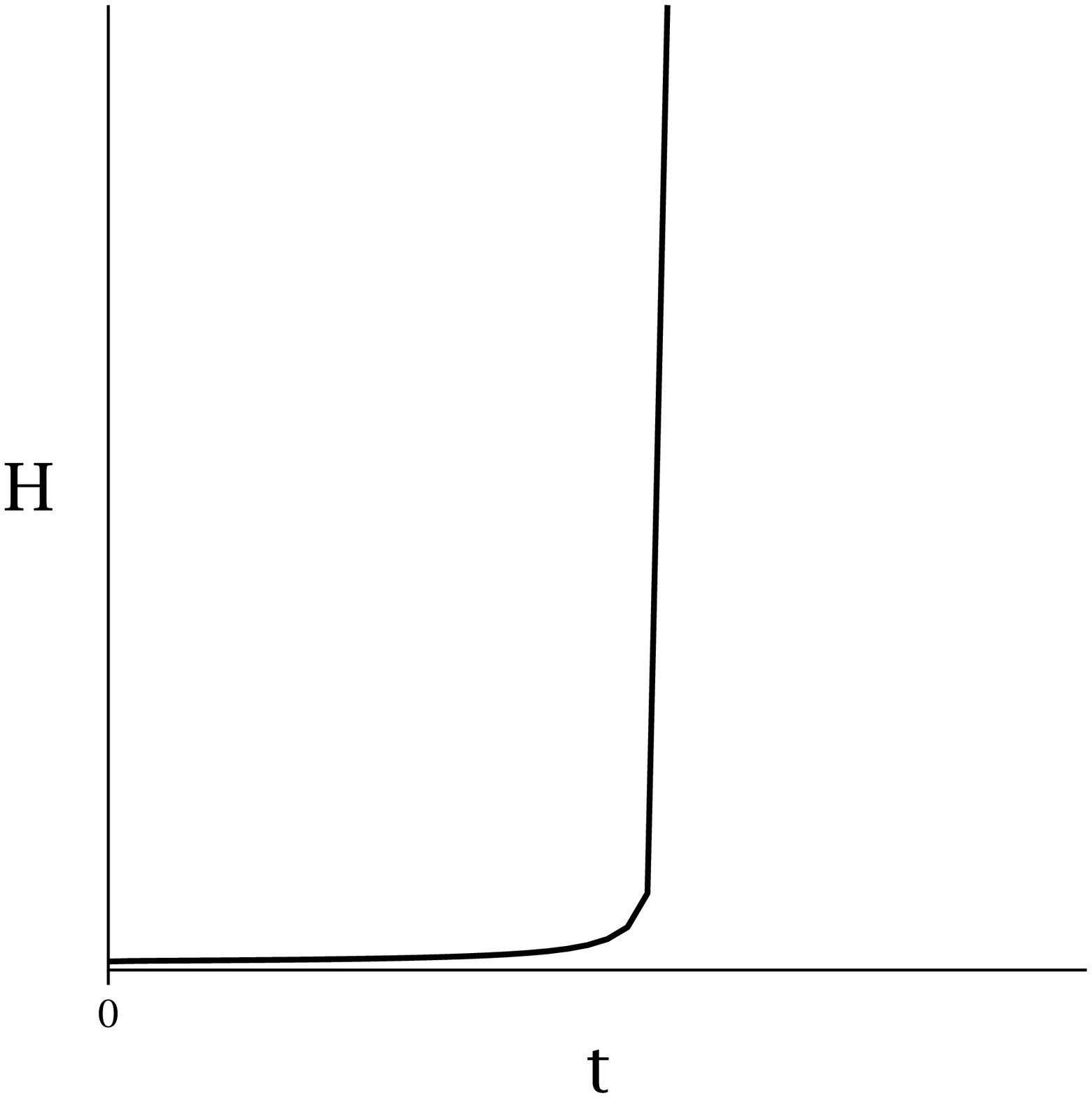}
\caption{Stable phantom solution for $f(R)=R+\alpha R^4$ does not depend on the coupling $\alpha$.}
\label{int:R4}
\end{minipage}
\hfill
\begin{minipage}[h!]{0.35\linewidth}
\includegraphics[width=1\linewidth]{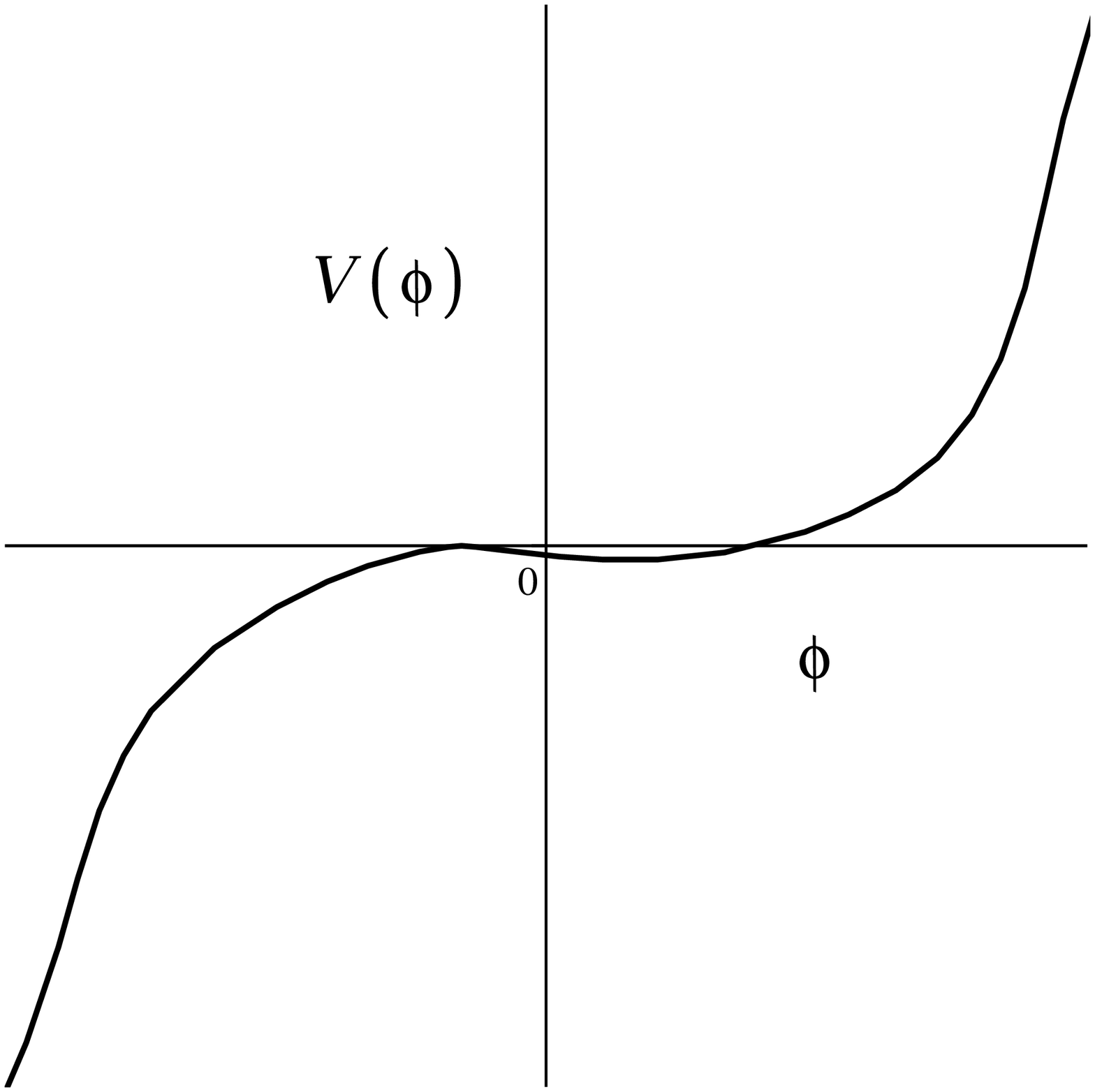}
\caption{Effective potential of scalaron for $f(R)=R-R^4$.}
\label{ris:R4}
\end{minipage}
\end{figure}

In this case:
\begin{align}
\label{v'r}
 V(\phi)& =\frac{\alpha}{3(2N-1)}\Bigl(R^N(2N^2-3N+1)-\alpha(N-2)(N-1)NR^{2N-1}\Bigr)\;,\\
\label{rn}
\phi & = \alpha NR^{N-1}-1\,. 
\end{align}
Two examples shown in Figs.\ref{ris:R+R4}, \ref{ris:R+R2} are 
\begin{align}
V(\phi)&=\frac{1}{12}(\phi^2+2\phi) \quad \textrm{for}\quad  f(R)=R+R^2\,,\\
\label{VR4}
\begin{split}
V(\phi)&=-\frac{1}{56}\sqrt[3]{2\phi+2}(2\phi-5)(\phi+1) \quad \textrm{for} \quad f(R)=R+R^4\,.
\end{split}
\end{align}

\begin{figure}[h!]
\begin{minipage}[h!]{0.35\linewidth}
\includegraphics[width=1\linewidth]{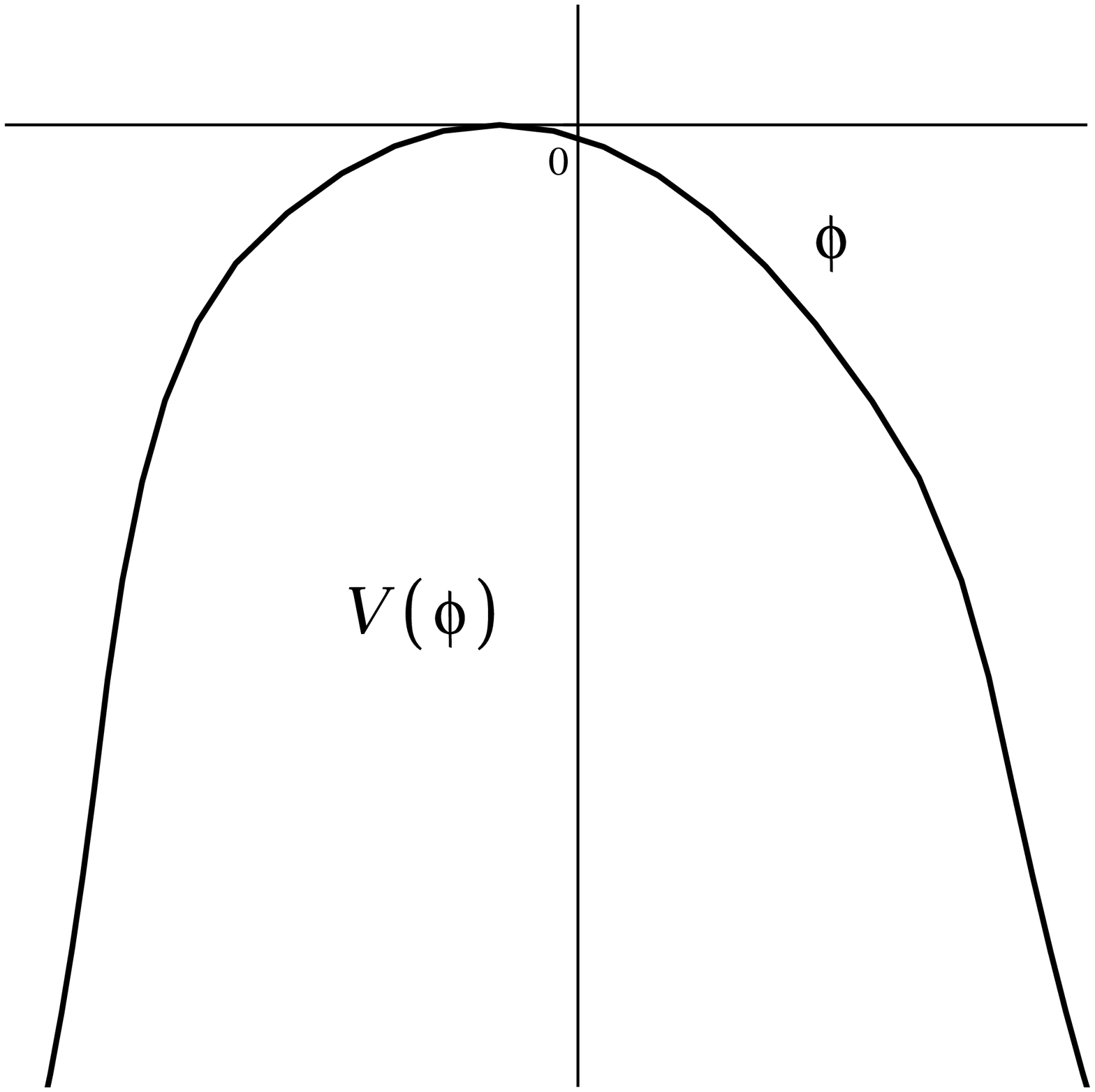}
\caption{Effective potential of scalaron for $f(R)=R-R^2$.} 
\label{ris:R2} 
\end{minipage}
\hfill
\label{int-R2} 
\begin{minipage}[h!]{0.35\linewidth}
\includegraphics[width=1\linewidth]{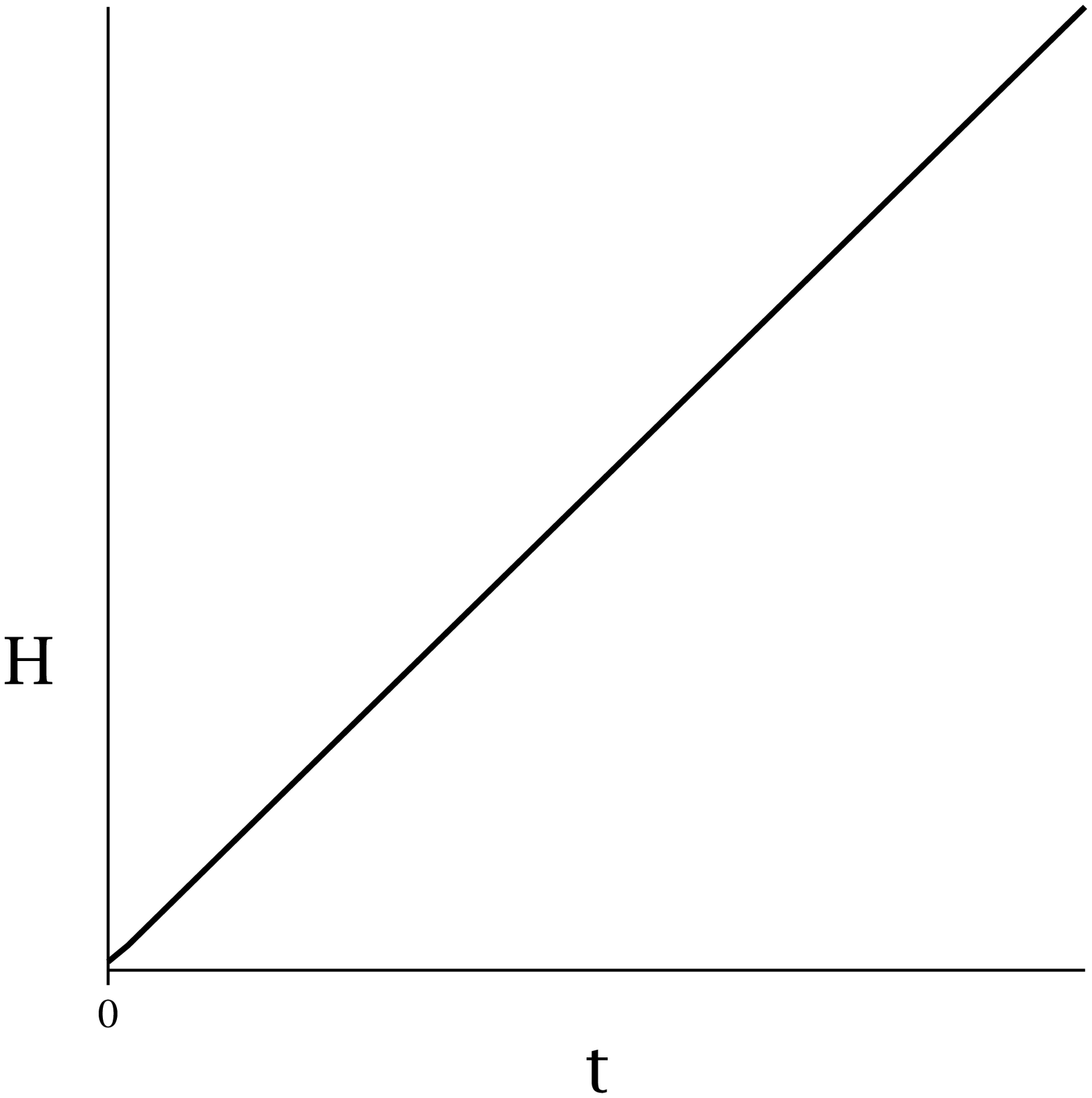}
\caption{Stable super-inflating behaviour for $f(R)=R-R^2$.} 
\label{int-R2}
\end{minipage}
\end{figure}

\begin{figure}[h!]
\begin{minipage}[h!]{0.35\linewidth}
\includegraphics[width=1\linewidth]{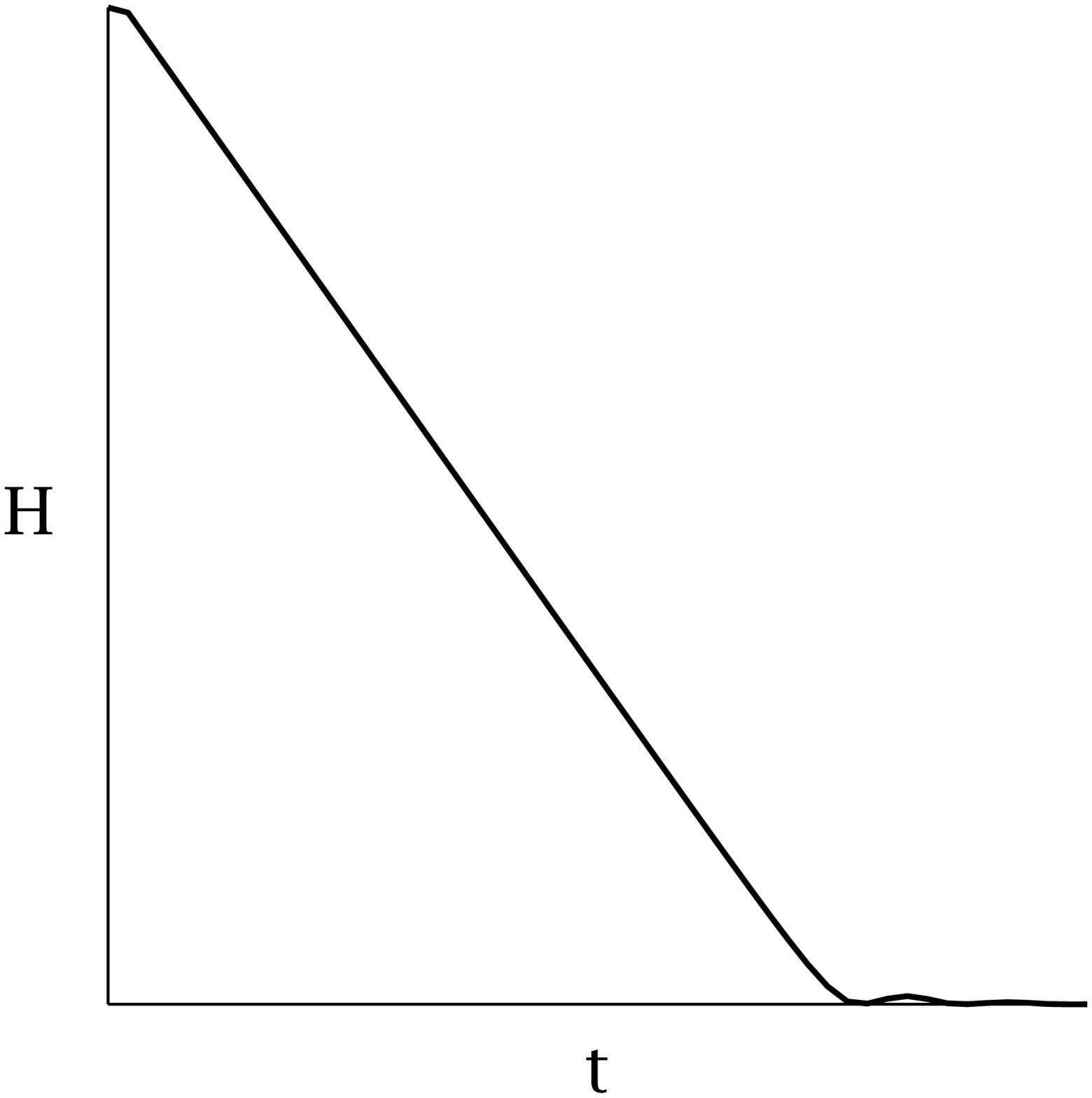}
\caption{Stable asymptotic solution for $f(R)=R+R^2$ leads to $H=0$.} 
\label{int:R2} 
\end{minipage}
\hfill
\begin{minipage}[h!]{0.35\linewidth}
\includegraphics[width=1\linewidth]{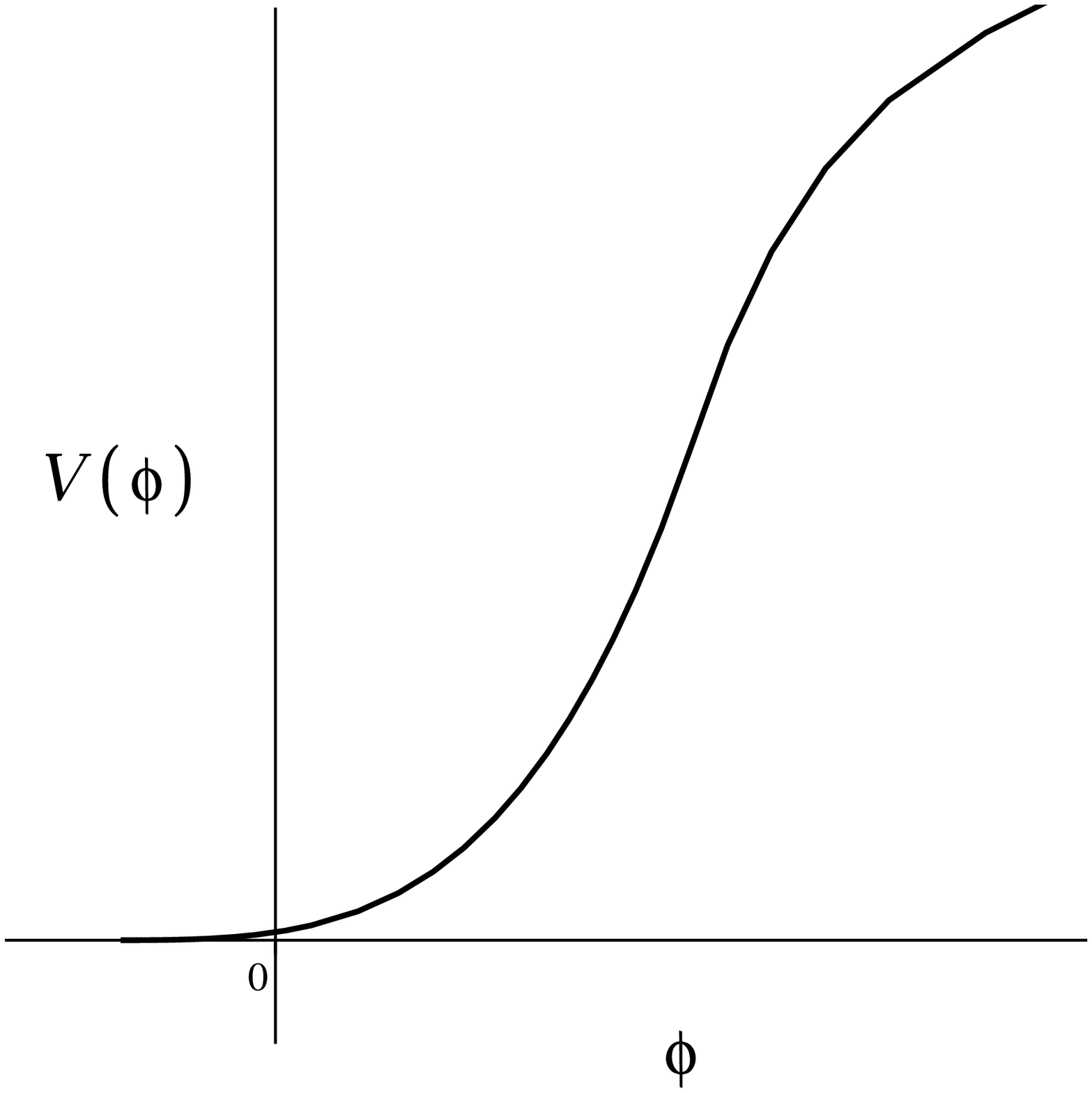}
\caption{Effective potential for $f(R)=R+R^{1.5}$.}
\label{graphR15}
\end{minipage}
\end{figure}

\paragraph{The case of N$>2$.}
For $N>2$ ($N=4$ on the figure, the local minimum is located at $R=0$) 
the potential has the asymptotics $V\to \infty$ at $R\to -\infty$ and $V\to-\infty$ at $R\to \infty$.
There is no global minimum and the "particle"  falls down into the region of large negative $V$. 
This  leads to $V\to-\infty$ 
and, correspondingly, $R\to\infty$\footnote{Consider theory with
odd $N$ we have to choose one of the branches ($R>0$ or $R<0$) because of the potential 
ambiguity. Also $R>0$ for theories with non-integer $N$.}. The numerical procedure confirms our reasoning (see Fig.\ref{int:R4}).
Hence in the case $N>2$ the curvature diverges and we obtain stable   
super-inflating solution. 
This solution appears to be a power-law phantom $a \sim t^\frac{2N^2-3N+1}{2-N}$
and have been found earlier in \cite{Schmidt}, \cite{COA} \cite{CDCT}.
Note, that the shape of the potential and asymptotic behavior does not depend essentially on the sign of $\alpha$ in this regime
(see Fig.\ref{ris:R4}).
\paragraph{The case of N$=2$.}
Effective potential of the scalar degree of freedom has the 
global minimum at $R=0$ for $\alpha>0$ (see Fig.\ref{ris:R+R2}) and the potential is unbounded from below and has 
asymptotic $V\to-\infty$ as $R\to\pm\infty$ for $\alpha<0$ (see Fig.\ref{ris:R2}).
Therefore super-inflation
(presented here by stable Ruzmaikin solution) exists in the last case (see Fig.\ref{int-R2}) and the case of positive $\alpha$ provides Minkowski solution (see Figs.\ref{int:R2}). The stability of Minkowski solution might be seen only numerically because $f'(0)\neq 0, f''(0)=0, f'''\equiv0$ in this case (see \cite{Muller}). Comparing with the previous case
we note that changing the sign
of the coupling constant in modified gravity with $R+\alpha R^2$ makes the super-inflation
behavior stable. On the contrary, it does not give new result for $R+\alpha R^N,\,N>2$.
\paragraph{The case of $1<$N$<2$.} 
By the definition, the curvature in such case must be positive. The effective potential is bounded from below for any couplings (see Fig.\ref{graphR15}). The numerical study shows that the theory is free from super-inflating solutions in such case. Detailed analysis of cosmological dynamics in this case was done in works \cite{CDCT,CTD} and it is out of topic of our work. Also we do not consider the case $N<1$ through the paper because it is sub-dominant in the high-curvature regime and, therefore, not essential for super-inflation. 

The results of the numerical procedure are in good agreement with those 
obtained by using the Dynamical system approach in \cite{CDCT},\cite{CTD}.	
In fact, we have only phantom stable power-law solutions for $N>2$.		

\subsection{f(R)=$R^N \exp(R)$} 
In this case:
\begin{align}
\begin{split}
V(\phi)&=-\frac{e^{2R}}{3}[N(N-1)(N-2)\sigma_{2N-2}+(3N^2-5N)\sigma_{2N-1}+(3N-2)\sigma_{2N}+\sigma_{2N+1}]\;,
\end{split}\\
\sigma_{k}&=\sum_{i=0}^{k}\frac{(-1)^{i}}{2^{i+1}}\frac{k!}{(k-i)!}R^{k-i}\,,\\
\phi & = (R^N+NR^{N-1})e^{R}-2 \,.
	\end{align}
\begin{figure}[h!]
		\begin{minipage}[h!]{0.35\linewidth}
			\includegraphics[width=1\linewidth]{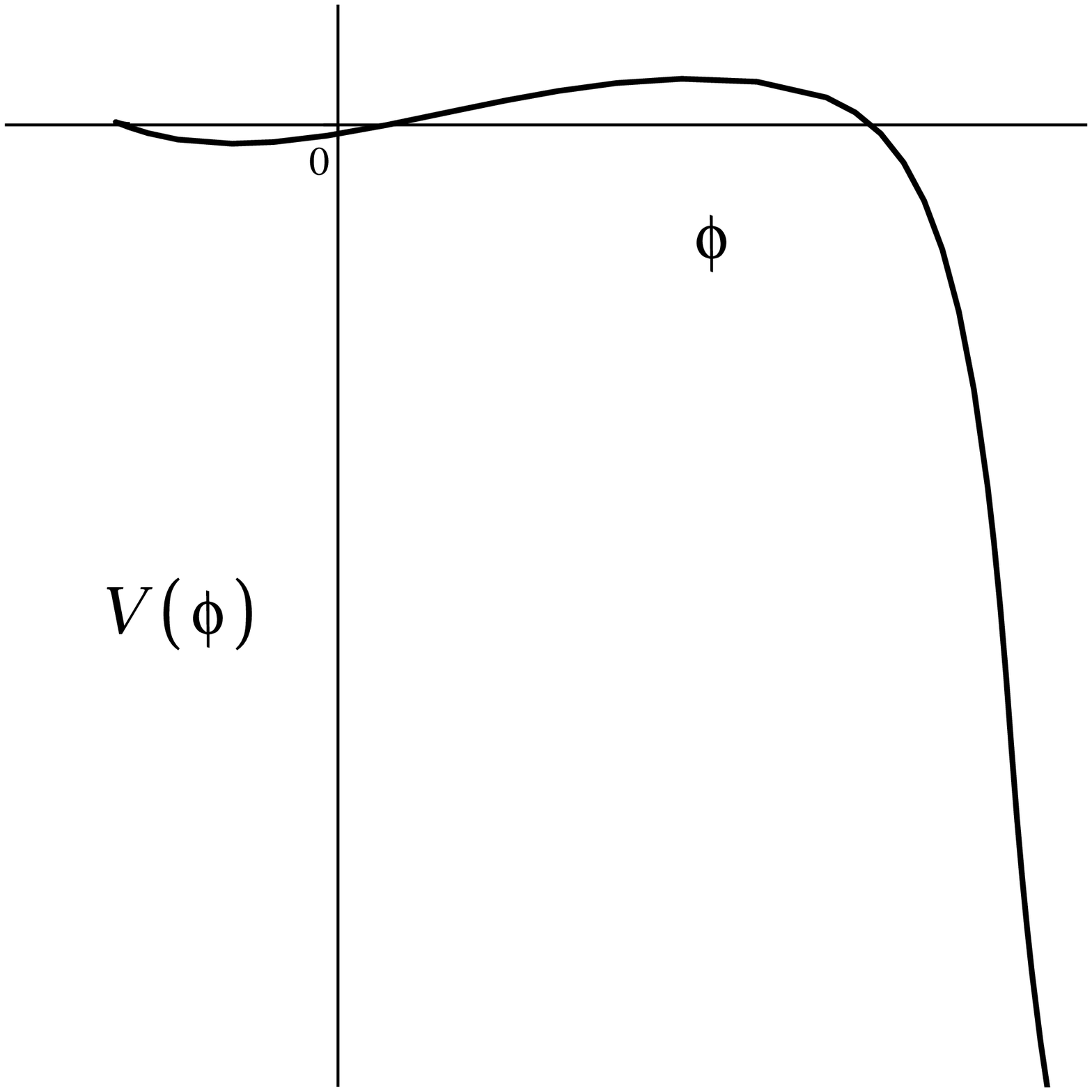}
			\caption{Effective potential of scalar degree of freedom for $f(R)=R^N\exp(R),\;1\leq N<2$.}  
			\label{ris:er}
				\end{minipage}
		\hfill
		\begin{minipage}[h!]{0.35\linewidth}
			\includegraphics[width=1\linewidth]{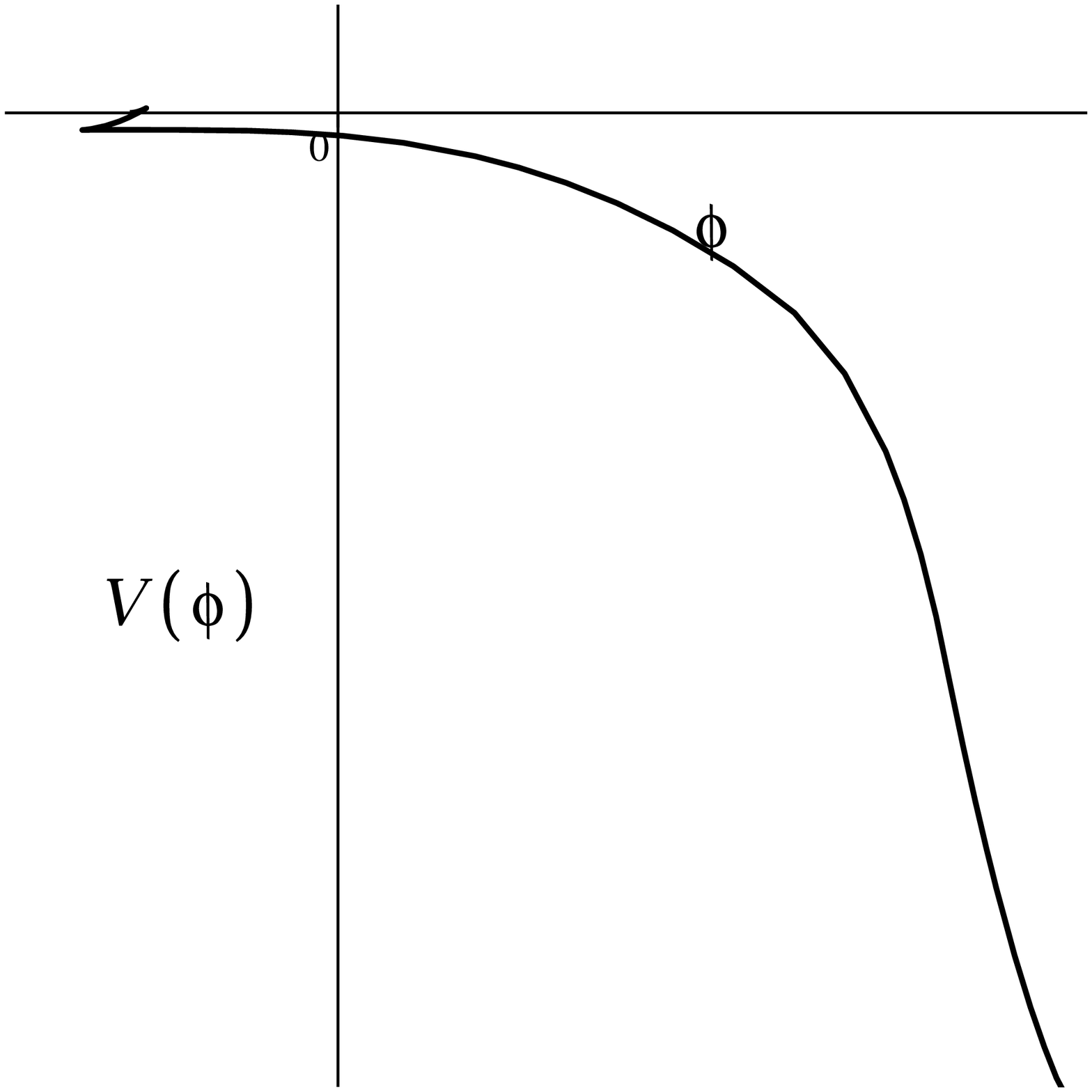}
			\caption{Effective potential of scalar degree of freedom for $f(R)=R^N\exp(R),\;N\geq 2$.}
			\label{ris:er2}
					\end{minipage}
\vfill
\begin{minipage}[h!]{0.35\linewidth}
\includegraphics[width=1\linewidth]{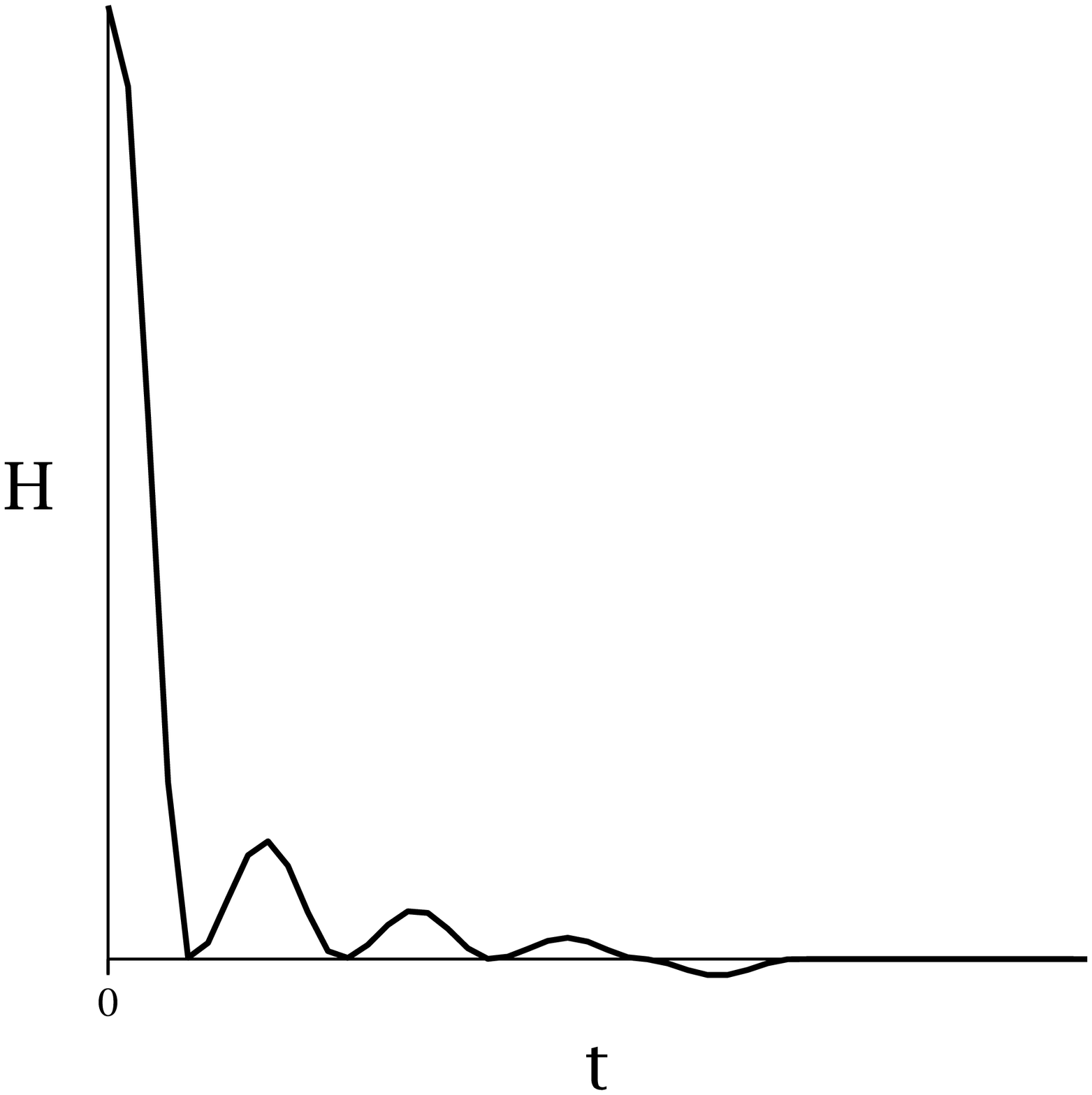}
\caption{Stable asymptotic solution for $f(R)=Re^R$ corresponding to the local minimum point at effective potential.}  
\label{int:er}
\end{minipage}
\hfill
\begin{minipage}[h!]{0.35\linewidth}
\includegraphics[width=1\linewidth]{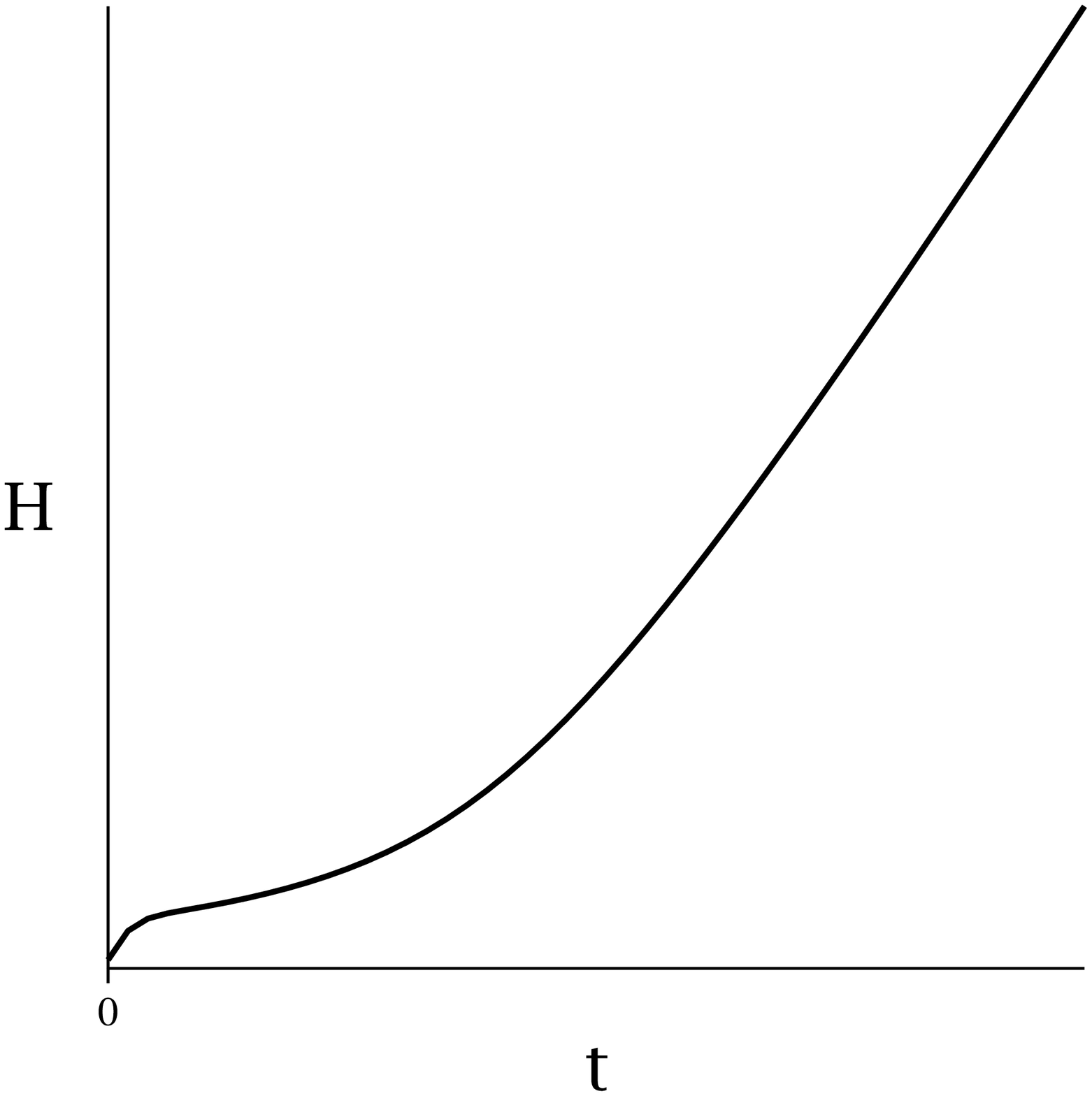}
\caption{Stable asymptotic super-inflation solution for $f(R)=R^Ne^R, N\geq1$.}
\label{int:er2}
\end{minipage}
\end{figure}	
\newpage
The potential has the asymptotic $V\to-\infty$ for $\phi \to \infty$ giving 
$R\to \infty$, hence the "particle" falls down in the potential hole (see Figs.\ref{ris:er}, \ref{ris:er2}). 
Moreover, the potential for  $f(R)=R^{N}\exp{(R)}, 1\leq N<2$ has the local minimum 
at $R=0$ and depending on initial conditions, one can obtain the stable Minkowski 
or super-inflating solutions.
The numerical procedure confirms our reasoning (see Figs.\ref{int:er}, \ref{int:er2}).\par 
Let us explain obtained results analytically. At first, consider de Sitter solutions in the theory:
\begin{align*}
f'(R_0)R_0-2f(R_0)=0\,,\Rightarrow \quad R_0=0 \quad \text{or} \quad R_0=2-N\,.
\end{align*}
Note that $R>0$ for dS in the chosen signature. Stability condition for de Sitter solution \cite{Muller} requires $0<f''(R_0)R_0<f'(R_0)$:
\begin{align}
\text{for $R_0=2-N$:} \quad  &\frac{f''(R_0)R_0}{f'(R_0)}=\frac{4-N}{2} \geq 1\,,\\
\text{for $R_0=0$:} \quad  &\frac{f''(R_0)R_0}{f'(R_0)}=\frac{N(N-1)}{N} < 1 \quad \text{for} \quad N<2\,.
\end{align}
Indeed, Minkowski solution is stable for $1\leq N<2$ and de Sitter solution is not.\par 
Secondly, diverging numerical solution demonstrates
linear behavior of the Hubble parameter hence one could expect that 
Ruzmaikin-like solution exists in such theory. Consider:
\begin{align}\begin{split}
a(t)&=a_0e^{\frac{Bt^{k+1}}{k+1}}\,,\\
H(t)&=Bt^k\,.
	\end{split}\end{align}
Substituting this ansatz in the equation of motion (\ref{numeq}) for $f(R)=R^N\exp{(R)}$, we find:
\begin{align}\begin{split}
& A_1t^{2k+2}+A_2t^2+A_3t^{3k+1}+A_4t^{k+1}+A_5t^{5k+3}+A_6t^{4k+2}+A_7t^{3k+3}+{}\\
&+A_8t^{k+3}+A_{9}t^{2k+4}+A_{10}t^{2k}+A_{11}t^{4k+4}=0\;.
\end{split}\end{align}
The biggest power indexes (which we should equate) belong to the terms $A_5t^{5k+3}$ and $A_{11}t^{4k+4}$.
This means that
\begin{align*}
4k+4&=5k+3\,,\\
k&=1\,.
	\end{align*}
Now we find the constant $B$. One should set the sum of coefficients in front 
of dominating term $A_{5}+A_{11}$ to zero:
\begin{align}
576kB^56^N-24B^46^N=0\,, \quad \Rightarrow \quad B=1/24\;.
\end{align}
Thus the Ruzmaikin-like solution $a\sim \exp{(t^2/48)}$ exists in 
the modified gravity theory with exponential correction for any $N\geq 1$. 
Note, that the Einstein term  in the case of $f(R)=R+\alpha R^{N}e^{R}$ is sub-dominant, and, therefore, is not necessary 
for Ruzmaikin solution here in contrast to the $R+R^2$ case (there is no such solution in a pure $R^2$ gravity). After neglecting
the Einstein term the constant $\alpha$ drops down from the equations of motion, and, therefore,  properties of the solution
does not depend on the sign of $\alpha$.   

This theory have been studied using dynamical system approach in \cite{CDCT}, though Ruzmaikin solution have not been found
there, so the list of asymptotic solutions from \cite{CDCT} is incomplete. 
  
\subsection{f(R)=$R+\alpha R^N \ln{R}$}
This function has some interesting features. We could expect that the logarithmic correction 
is not essential for high-curvature regime, 
and this theory would be equivalent to $f(R)=R+\alpha R^N$ described above in this paper. However,this correction 
allow us to obtain new solutions for $N=1,2$ which do not exist without them. It is worth to note that 
despite logarithm diverges at zero, the chosen form of $f(R)$ ensures that it is regular in this point. What can 
happen for a singular $f(R)$, we discuss in the next section.

\begin{figure}[h!]
			\begin{minipage}[h!]{0.35\linewidth}
			\includegraphics[width=1\linewidth]{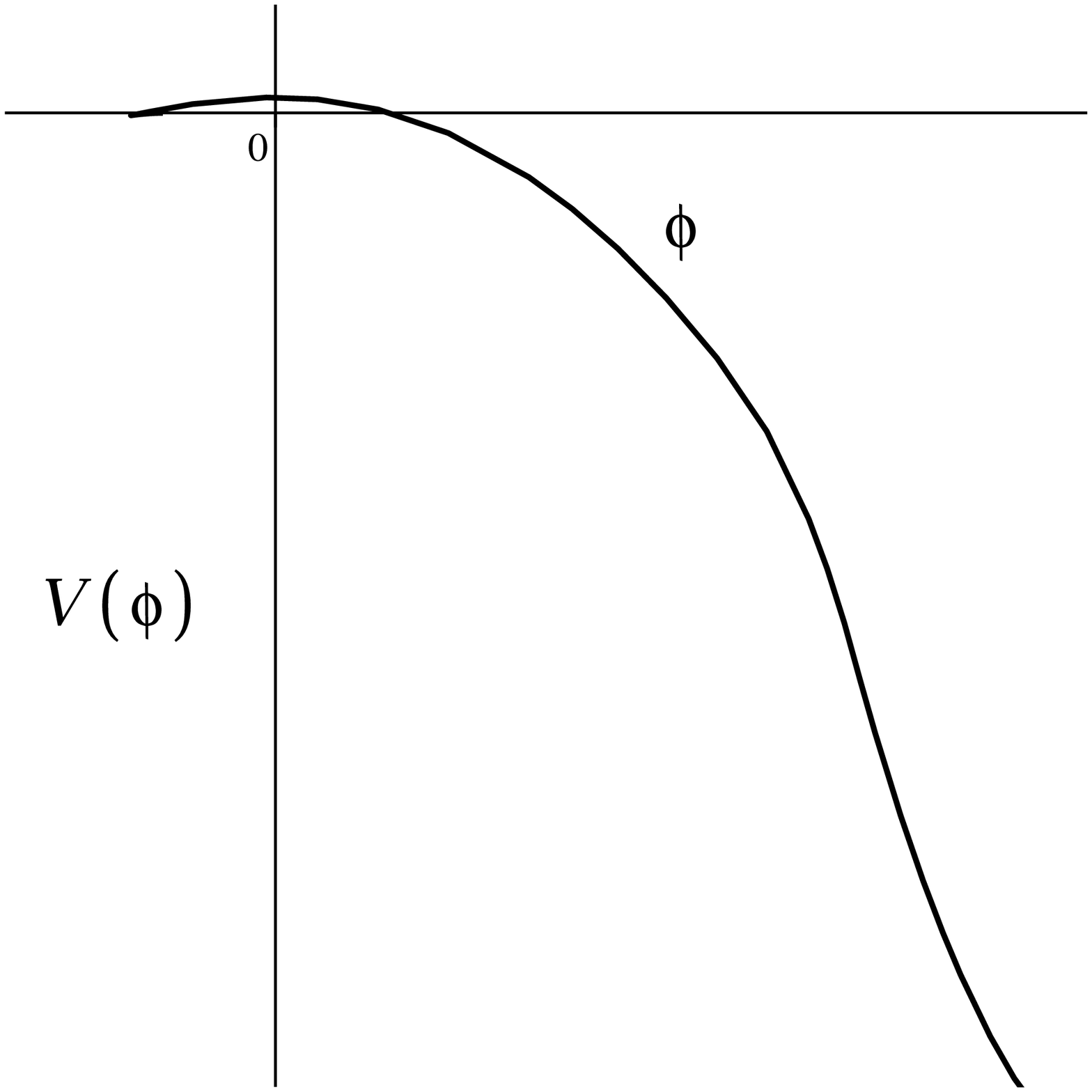}
			\caption{Effective potential for $f(R)=R+R^N\ln{R}, \, N>2$.}  
						\label{ris:rnlnr}
				\end{minipage}
		\hfill
		\begin{minipage}[h!]{0.35\linewidth}
			\includegraphics[width=1\linewidth]{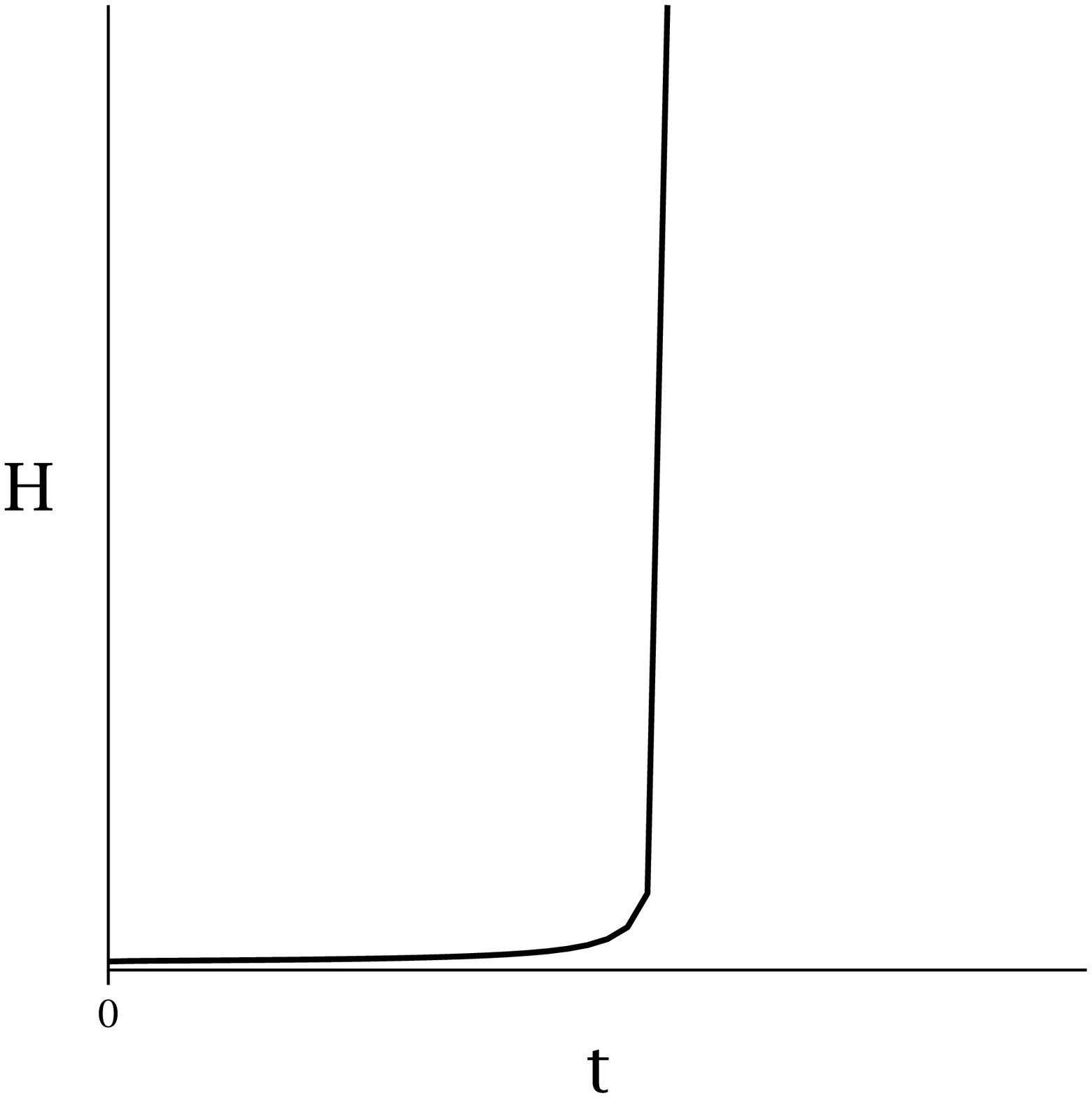}
			\caption{Stable phantom solution for $f(R)=R+R^N\ln{R}$, $N>2$.}
			\label{int:rnlnr}
					\end{minipage}
			\end{figure}

Let us write down the equations for scalar field and its potential in the terms of the $R$ as usual.\\
\begin{align}\begin{split}
V(\phi)=&-\,{\frac {\alpha}{ 3\left( 2\,N-1
 \right) ^{3}}}\Biggl[\left( -16\,{N}^{4}-11\,{N}^{2}
+2\,N +4\,{N}^{5}+21\,{N}^{3}
 \right)\alpha {R}^{2\,N-1}  \left( \ln R \right) ^{2}{}\\ 
& + (21\,{N}^{2}-
10\,N+2\,+8\,{N}^{4}-22\,{N}^{3})\alpha {R}^{2\,N-1}\ln R {}\\
& + (20\,{N}^{3}-1-8\,{N}^{4}-18\,{N}^{2}+7\,
N)\alpha {R}^{N}  \ln R \\
&+(4\,{N}^{3}-3\,{N}^{2}+1)\alpha {R}^{2\,N-1}+(1+12\,{N}^{2}-8\,{N}^{3}-6\,N){R}^{N}\Biggr]\;,
\end{split}\\
\phi  = &\alpha(N\ln{R}+1){R}^{N-1}-1 \,.
\end{align}
For example in the case of $N=1, \alpha=1$ these expressions reduce to:
\begin{equation}
 V(\phi) =\frac{1}{3}R(\ln{R}-1),\quad \phi  = \ln{R}\;.
	\end{equation}
\paragraph{The case of N$>2$.} 
For all $N>2$ the effective potential has no global minimum. 
Thus this theory should contain regimes with diverging curvature. Numerical calculation shows stable phantom solution in the considered case (see Figs.\ref{ris:rnlnr} and \ref{int:rnlnr}).
The form of phantom solutions found for $f(R)=R+\alpha R^N$ 
in \cite{CDCT} does not
change (it is easy to see using power-law ansatz), which is natural to expect due to very slow growth
 of the logarithm function:
 \begin{equation}
 \label{Phantom}
 a(t)=a_0 (t-t_0)^{\frac{2N^2-3N+1}{2-N}}\,.
 \end{equation}
Note that the numerical solution does not depend on the 
coupling in front of logarithmic term as well as the shape of effective potential
in this case.

\begin{figure}[h!]

\begin{minipage}[h!]{0.32\linewidth}
\includegraphics[width=1\linewidth]{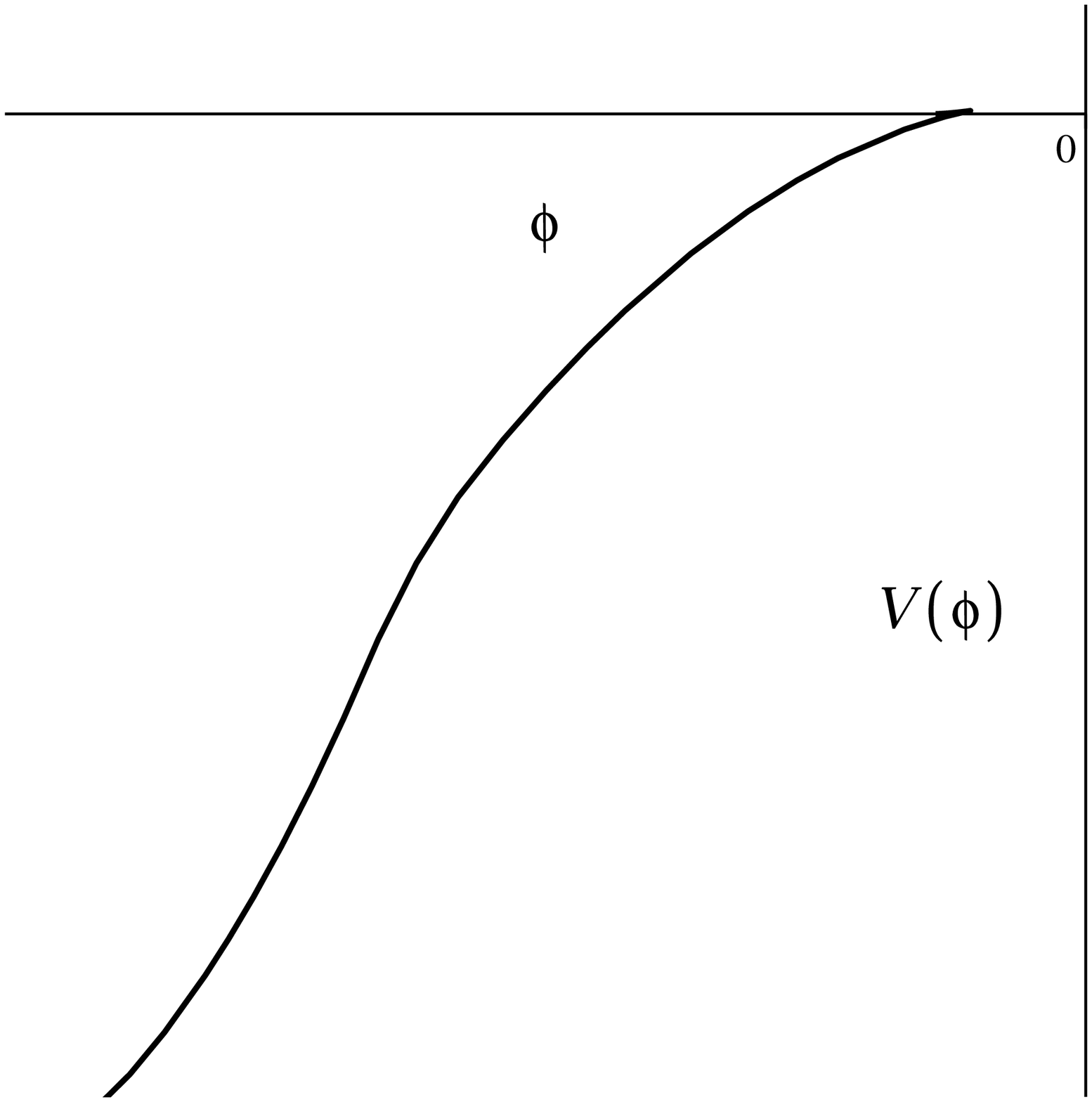}
\caption{Effective potential for $f(R)=R-R^N\ln{R}$, \mbox{$N\geq 2$}.}
\label{ris:r-rnlnr}
\end{minipage}
\hfill
\begin{minipage}[h!]{0.32\linewidth}
\includegraphics[width=1\linewidth]{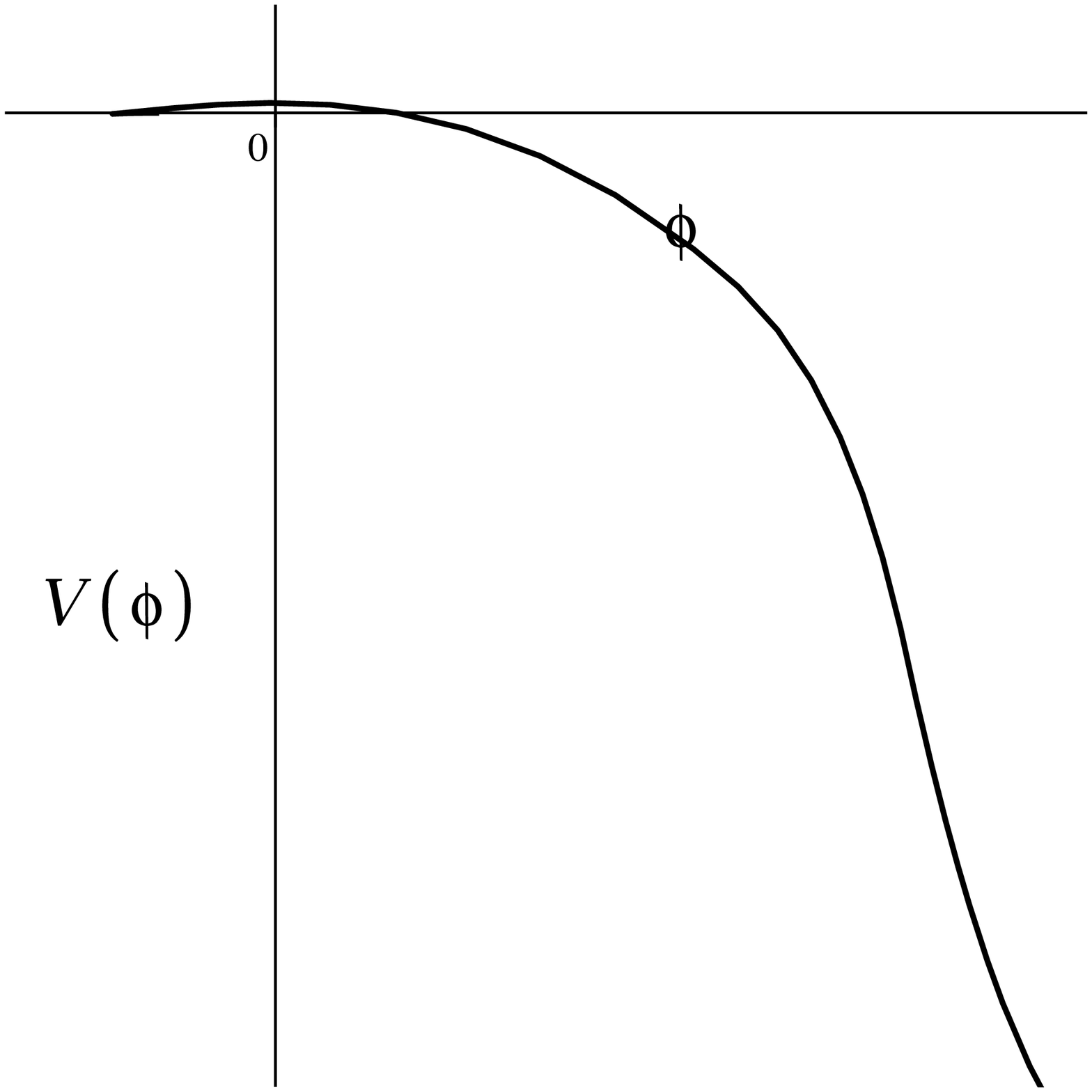}
\caption{Effective potential for \mbox{$f(R)=R+R^2\ln{R}$}.}
\label{ris:r+r2lnr}
\end{minipage}
\vfill
\begin{minipage}[h!]{0.33\linewidth}
\includegraphics[width=1\linewidth]{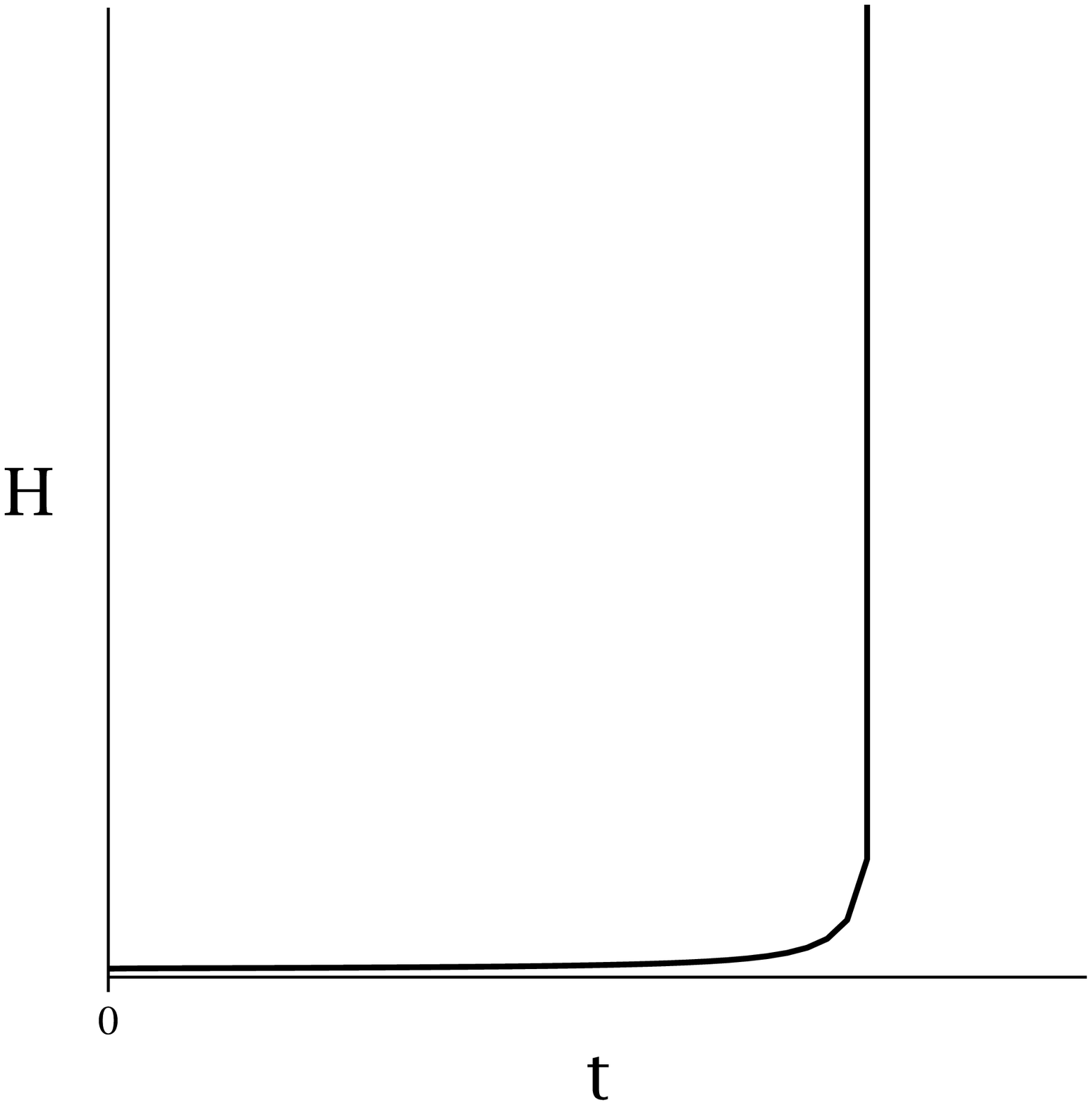}
\caption{Stable super-inflating solution for $f(R)=R+R^2\ln{R}$.}  
\label{int:r-r2lnr}
\end{minipage}
\hfill
		\begin{minipage}[h!]{0.33\linewidth}
			\includegraphics[width=1\linewidth]{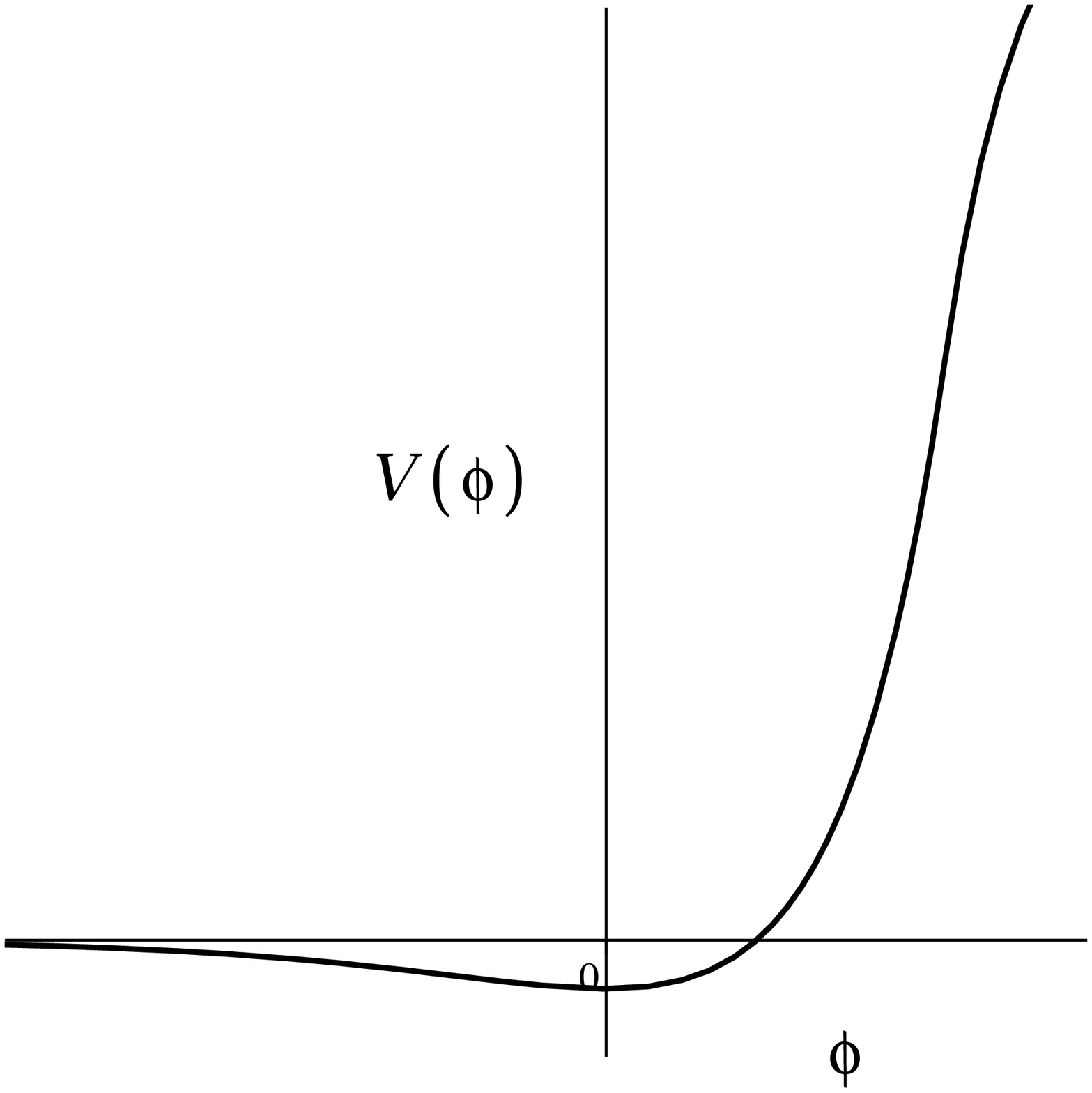}
			\caption{Effective potential for $f(R)=R+R\ln{R}$.}  
			\label{ris:rlnr}
				\end{minipage}
\vfill
		\begin{minipage}[h!]{0.33\linewidth}
			\includegraphics[width=1\linewidth]{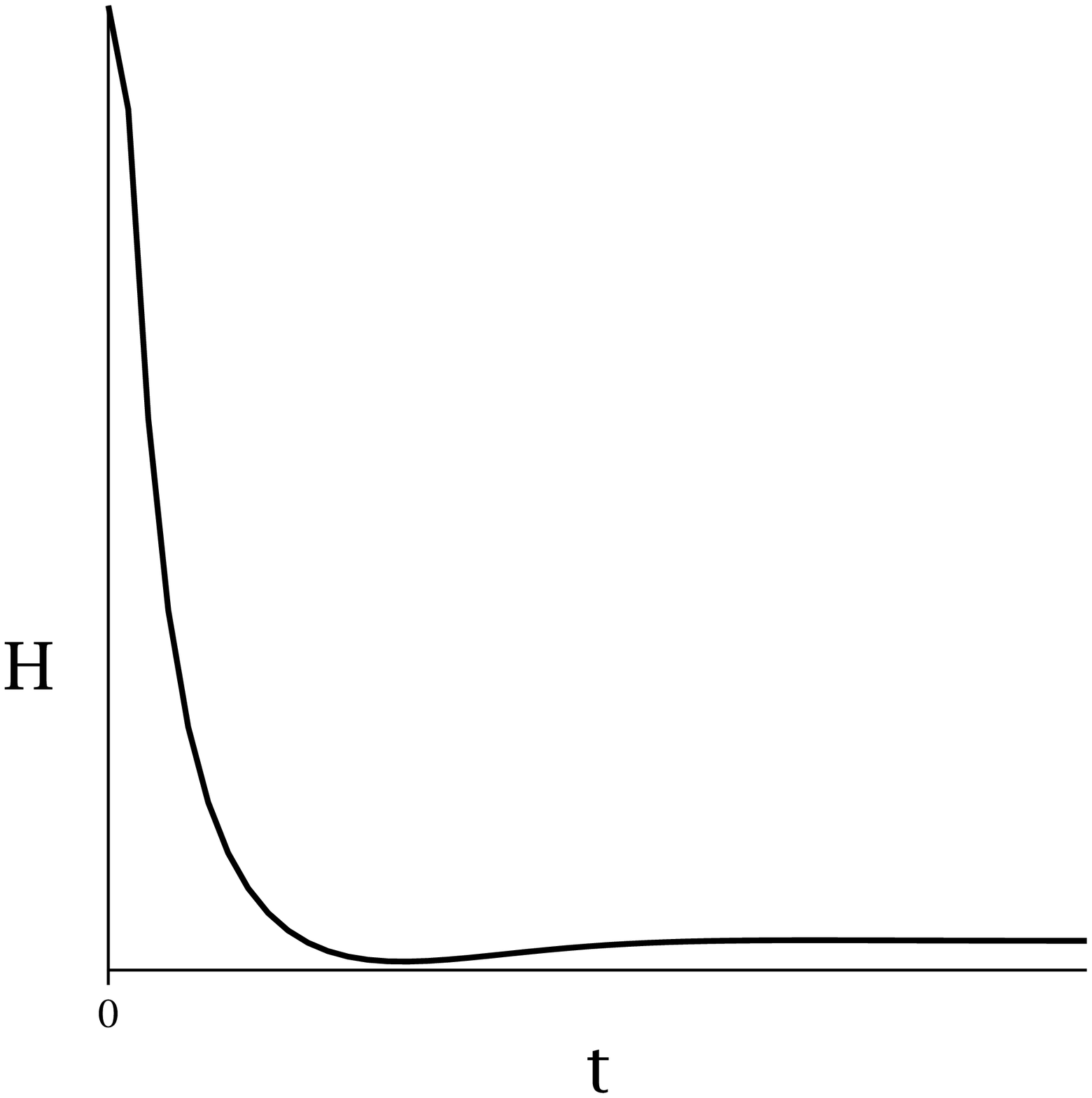}
			\caption{Stable de Sitter solution for $f(R)=R+R\ln{R}$.}
			\label{int:rlnr}
					\end{minipage}
		\hfill
\begin{minipage}[h!]{0.33\linewidth}
\includegraphics[width=1\linewidth]{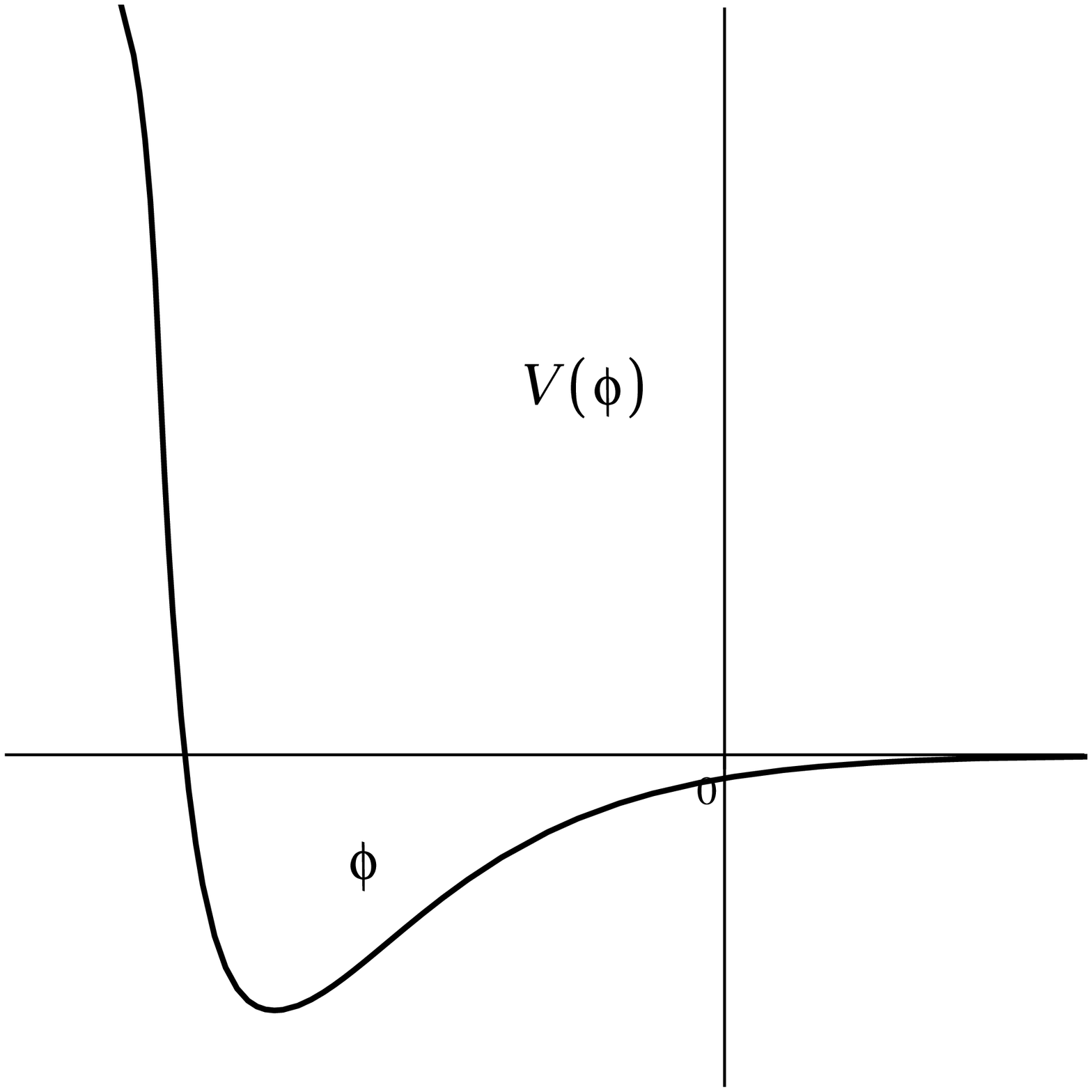}
\caption{Effective potential for $f(R)=R-R\ln{R}$.}  
\label{ris:r-rlnr}
\end{minipage}
			\end{figure}
\par


\paragraph{The case of N$=2$.} We remind  a reader that for $f(R)=R+\alpha R^N$ the case of $N=2$ only
gives us the Ruzmaikin solution (even if we allow non-integer power indexes). Effective potentials in the theory with
logarithmic correction
for $N=2$ and $N>2$ do not depend much on coupling $\alpha$ and 
also do not significantly differ from each other
(see Figs.\ref{ris:rnlnr}, \ref{ris:r-rnlnr} and \ref{ris:r+r2lnr}). All such potentials are unbounded from below, however, 
the Ruzmaikin solution existing in the theory without logarithmic correction, disappears when we take this corrections
into account. 
Choosing the ansatz in the form $H=Bt^k$, we obtain for $f(R)=R+\alpha R^2\ln R$:
\begin{align}\begin{split}
t^2+ A_1t^{k+1}\ln{(12B^2t^{2k}+6Bkt^{k-1})}+A_{2}t^{2k+2}+
A_{3}\ln{(12B^2t^{2k}+6Bkt^{k-1})}+A_{4}t^{k+1}+A_{5}=0\;.
\end{split}\end{align}
The first term originates from the Einstein term. The third term becomes 
dominant for large times and it can not be compensated by another terms. 
Therefore, there is no Ruzmaikin solution in this theory. Using the power-law 
ansatz $a(t)=a_0 (t-t_0)^{A}$ one can easily obtain that there is no simple power-law phantom solution too. 
However, our numerical procedure shows the super-inflating solution 
diverging at a finite time as presented in Fig. 
\ref{int:r-r2lnr}. 
\par
Let us explain the nature of such solution 
in the theory with logarithmic correction.
Consider:
\begin{align}
\label{Hln}
\begin{split}
H(t)&=B\frac{\ln[\Lambda(t_0-t)]}{t_0-t}=B\frac{\ln\Lambda+\ln(t_0-t)}{t_0-t}\;,
	\end{split}\end{align}
Substituting this ansatz in the equation of motion (\ref{numeq}) we find:
\begin{align}\begin{split}
& -\alpha^{-1}(t_0-t)^2\ln^2[\Lambda(t_0-t)]+42\ln[\Lambda(t_0-t)]+(54B-30)\ln^2[\Lambda(t_0-t)]-54B\ln^3[\Lambda(t_0-t)]\\
& +12B^2\ln^4[\Lambda(t_0-t)]+6\ln(R)+24\ln(R)\ln[\Lambda(t_0-t)]+(36B-18)\ln(R)\ln^2[\Lambda(t_0-t)]\\
&-36B\ln(R)\ln^3[\Lambda(t_0-t)]+6=0\;.
\end{split}\end{align}
In the high-curvature case
\[
R=\frac{12B^2\ln^2[\Lambda(t_0-t)]}{(t_0-t)^2}-\frac{6B}{(t_0-t)^2}+\frac{6B\ln[\Lambda(t_0-t)]}{(t_0-t)^2}\;,
\]
\begin{equation}
\label{hcl}
\ln(R)\approx -2\ln(t_0-t)+\mathcal{O}(\ln\ln\mid t_0-t\mid)\,,
\end{equation}
so terms with $12B^2$ and $-36B$ factors dominate ($t\to t_0$) and 
lead to approximate solution
\mbox{$H\sim \ln[\Lambda(t_0-t)]/(t_0-t)$}, 
which have been found numerically. 
The $\Lambda$ constant determines the time 
scale for which this solution exists and depends on the initial conditions.
To obtain the constant $B$ one should set the coefficient in front 
of dominating terms to zero (factor $-2$ arises from (\ref{hcl})):
\begin{equation}
12B^2+72B=0 \quad \Rightarrow B=-6\;.
\end{equation}
The Einstein 
term is not important for this regime in such theory as well as in the previous subsection,
thus stable super-inflating solution exists for
any $\alpha \neq 0$.\par

\paragraph{The case of $1\leq$ N$<2$}

The potential has a global minimum (at $\phi=0$, $R=1$ in the case $\alpha=1,N=1$), 
flat small curvature asymptotic $\phi\to-\infty$ corresponding to  $R\to 0$ and exponential potential 
wall for large curvature (see Fig.\ref{ris:rlnr}). Thus this theory does not contain  solutions with diverging curvature. The numerical procedure demonstrates stable de Sitter regime (see Fig.\ref{int:rlnr}) in the case \mbox{$1\leq N<2$} for each certain coupling $\alpha$ even for $N$ close to $2$. Effective potentials
in this case does not depend significantly on the 
sign in front of logarithmic term (see Figs.\ref{ris:rlnr} and \ref{ris:r-rlnr}).
The existence and stability of de Sitter solution (as well as the absence of super-inflating solutions) for $N=1$ have been proved 
using the dynamical systems methods in \cite{Frolov}.\par 
Let us study the existence and stability of de Sitter solution in the case $N=2-\epsilon, \epsilon\ll 1$:
\begin{align}
f'(R)R-2f(R)&=0\,,\Rightarrow \\
R+\alpha(N\ln(R)+1)R^N-2R-2\alpha R^N\ln(R)&=0\,,\\
-R+\alpha R^N[(N\ln(R)+1)-2\ln(R)]&=0\,,
\end{align}
The numerical procedure shows existence of a stable de Sitter solution with large curvature  $R\gg1$. Thus the Einstein term is negligible for $\epsilon\ll 1$ and $\alpha$ drops out from the equations of motion: 
\begin{align}
\alpha R^N[(N-2)\ln(R)+1]=0\,,\\
\label{NPdS}
R= e^{1/(2-N)}\,.
\end{align}
Stability condition $0<f''(R_0)R_0<f'(R_0)$ (see \cite{Muller}) in the considered case implies:
\begin{align}
\label{StabDS}
2-\sqrt{2}<N<2\,.
\end{align}
Indeed, in the theory with $R+\alpha R^{2-\epsilon}\ln R$ we have no super-inflating solutions and obtain stable de Sitter solution with $R_0\approx e^{{1}/{\epsilon}}$. 
Note that both solutions ((\ref{Phantom}) and (\ref{NPdS})) with $N=2\pm \epsilon$ are sufficiently non-perturbative and limiting $\epsilon \to 0$ does not allow us obtain solution existing only for $N=2$. The situation here is the same with the known case $R+\alpha R^{2+\epsilon}$, where Ruzmaikin solution exists only for $\epsilon=0$ and does not exist for any non-zero $\epsilon$.\par 
As in the former case, the de Sitter solution is stable, the only important difference is that the existence and stability of de Sitter
solution with the logarithmic corrections does not depend upon the sign of the coupling constant $\alpha$ in the limit 
of small $\epsilon$. In the general case of an arbitrary $N \in [1,2]$ the situation depends on both $N$ and $\alpha$,
detailed description of de Sitter solution for this case is beyond the scope of this paper.
 
\subsection{f(R)=$R-1/R$ }
\begin{figure}[h!]
		\begin{minipage}[h!]{0.35\linewidth}
			\includegraphics[width=1\linewidth]{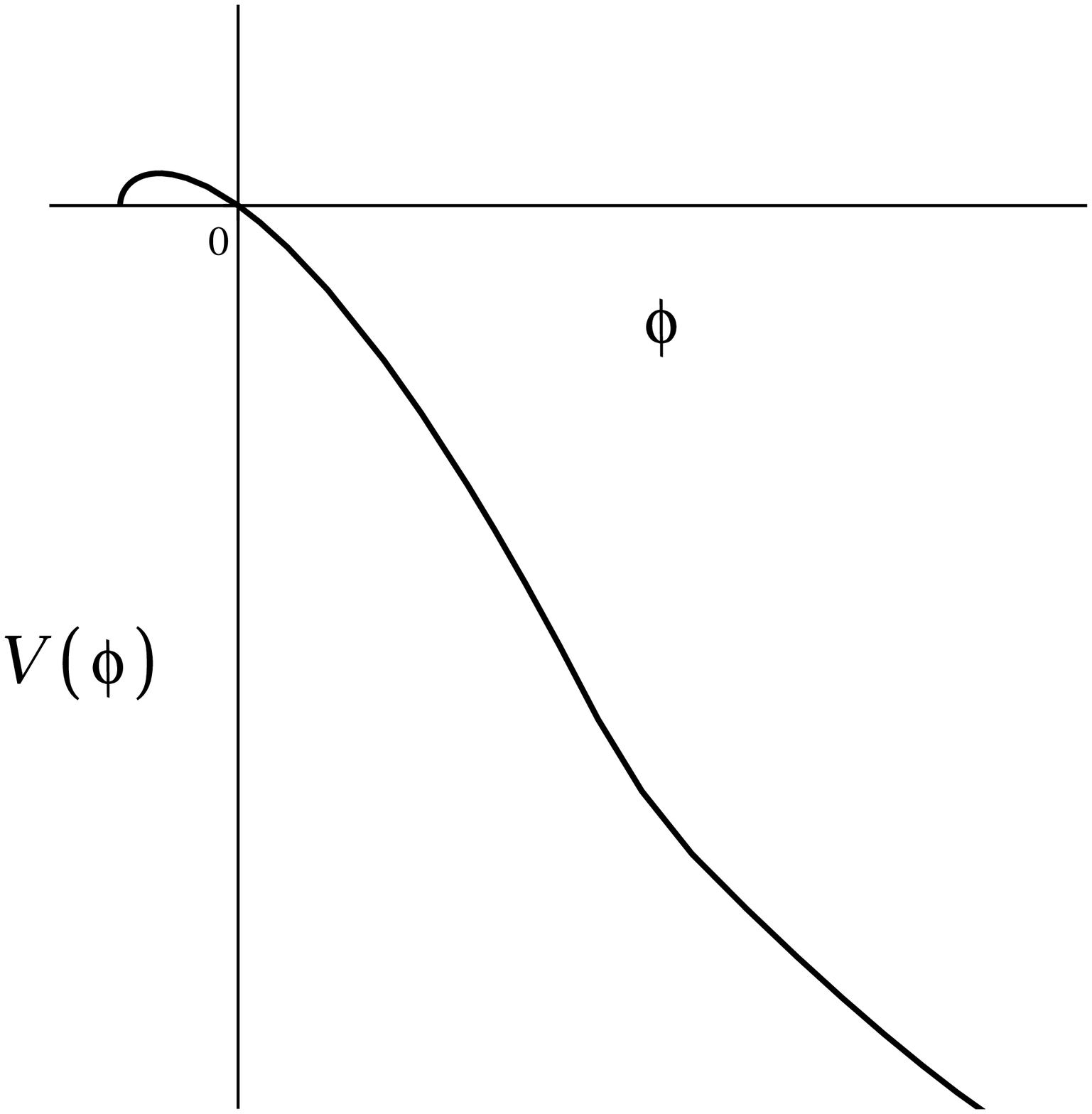}
			\caption{Effective potential for $f(R)=R-1/R, \,R>0$.} 
					\label{ris:1/r} 
				\end{minipage}
		\hfill
		\begin{minipage}[h!]{0.35\linewidth}
			\includegraphics[width=1\linewidth]{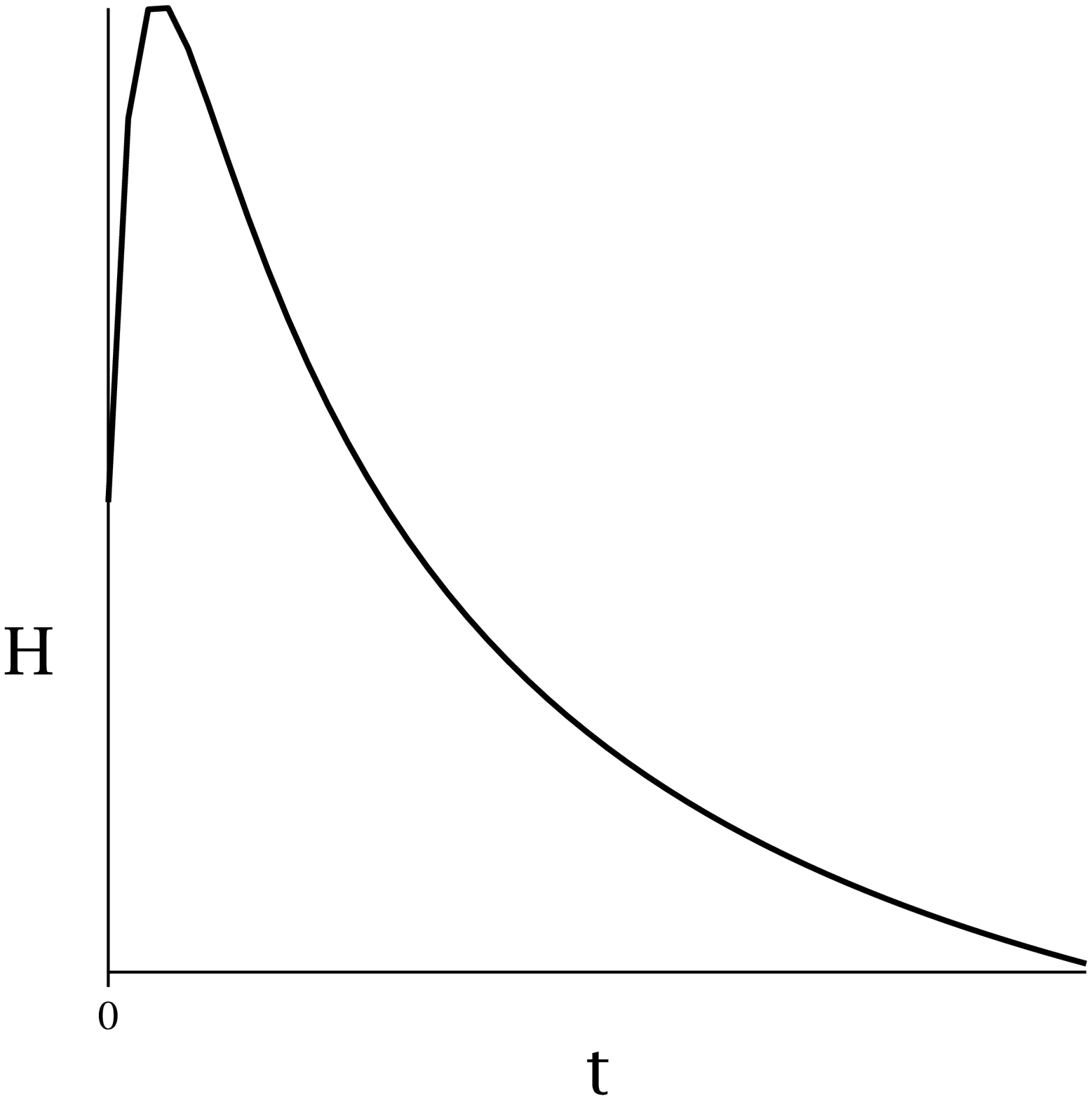}
			\caption{Stable numerical solution for $f(R)=R-1/R$.}
			\label{int:1/r}
					\end{minipage}
			\end{figure}

In this subsection we discuss that can happen if $f(R)$ and (or) $V(\phi)$ 
is singular for the finite value of its argument. In this case 
unboundedness of the effective potential from below
does not indicate a super-inflating solution because $R$ can 
tend to a finite limiting value for which $f(R)$ diverges.
Consider the prototype of General Relativity infra-red modification, proposed in \cite{CC}. 
It has been already studied that the scalar degree of freedom is unstable in this 
theory (\cite{Dolgov}), but it is illustrative example in which mechanical analogue
still works and gives results different from obtained above.  
The scalar field for this model is:
\begin{align}
V(\phi)&=\frac{2}{3}\,{\frac {{R}^{2}-1}{{R}^{3}}}\;,\\
\phi &=\frac{1}{R^2}-1\;.
\end{align}
$V(\phi)$ is an ambiguous function of $\phi$, thus consider the branch $R>0$  
(see Figs.\ref{ris:1/r} and \ref{int:1/r}).
			\\
The potential has the asymptotic $V\to-\infty$ for 
$\phi \to \infty$ corresponding to  $R\to 0$ , so the "particle" 
falls down in the potential hole and provides a stable Minkowski solution\footnote{
The stability was checked only numerically because $m^2=0$ (see \cite{Muller}) in this case.}. 
As a result, the potential unbounded from below does not lead 
to a super-inflating
solution in this case.  
\newpage
\section{Conclusion}
It this paper we have explored the idea of determining asymptotic cosmological behavior in $f(R)$-gravity using the effective potential of scalar degree of freedom for regular functions $f(R)$. The presence of the global minimum appears to be incompatible  with super-inflating cosmological solutions, on the other hand,
unboundedness of the potential from below indicates existence of such solutions. We considered only vacuum case; in the presence of matter the  dynamics is not so evident.\par

As an application of this property we considered three families of functions $f(R)$ very popular in current researches on modified
gravity, namely $R+\alpha R^N$, $R+\alpha R^N \exp (R)$ and \mbox{$R+\alpha R^N \ln R$}. 
It appears that only the following potentials from these
infinite families are free from stable super-inflation solution: the known $R+\alpha R^N, 1<N<2$; $R+\alpha R^2, \alpha>0$ cases and the case of $R+\alpha R^{N}\ln R\,,1\leq N<2$.
 All other $f(R)$ functions from these families contain stable super-inflating solutions, some of them have not
been discovered earlier. 

\section*{Acknowledgments}
Authors would like to thank Andrei Frolov for stimulating discussions. This work was supported by Federal Russian Science Agency through the research contract 02.740.11.0575 and by RFBR grant 11-02-00643.

\end{document}